\documentclass[AMA,Times1COL]{WileyNJDv5} 

\articletype{RESEARCH ARTICLE}%

\received{Date Month Year}
\revised{Date Month Year}
\accepted{Date Month Year}
\journal{Journal}
\volume{00}
\copyyear{2023}
\startpage{1}

\usepackage[T1]{fontenc}
\usepackage{geometry}
\usepackage{amsmath}

\usepackage{amsthm}
\usepackage{stmaryrd}
\usepackage{graphicx}
\usepackage{booktabs}
\usepackage{multirow}
\usepackage{makecell}
\usepackage{xurl}
\usepackage{esint}
\usepackage{algorithm}
\usepackage{algpseudocode}
\usepackage{adjustbox}
\makeatletter

\providecommand{\proofname}{Proof}
\fi

\journal{submitted to IJNME}

\usepackage{hyperref}
\hypersetup{colorlinks = true, allcolors = blue}

\usepackage[nameinlink]{cleveref}

\crefname{figure}{Fig.}{Figs.}
\crefformat{equation}{Eq.~#2(#1)#3}
\crefformat{section}{Section~#2#1#3}
\AtBeginDocument{%
	\let\cite\cite
}

\usepackage[labelfont=bf]{caption}
\@ifundefined{showcaptionsetup}{}{%
	\PassOptionsToPackage{caption=false}{subfig}}
\usepackage{subfig}
\makeatother

\usepackage{babel}

\begin{document}

\title{A Pretraining-Finetuning Computational Framework for Material Homogenization}

\author[rvt,rvt3]{Yizheng Wang}


\author[rvt2]{Xiang Li}

\author[rvt]{Ziming Yan}

\author[rvt]{Shuaifeng Ma}

\author[rvt,rvt4]{Jinshuai Bai}

\author[rvt7]{Bokai Liu}

\author[rvt6]{Xiaoying Zhuang}

\author[rvt3]{Timon Rabczuk}

\author[rvt]{Yinghua Liu}

\address[rvt]{\orgdiv{Department of Engineering Mechanics}, \orgname{Tsinghua University}, \orgaddress{\state{Beijing}, \country{China}}}

\address[rvt2]{\orgdiv{School of Information Science and Technology}, \orgname{Hainan Normal University}, \orgaddress{\state{HaiKou}, \country{China}}}

\address[rvt3]{\orgdiv{Institute of Structural Mechanics}, \orgname{Bauhaus-Universit\"{a}t Weimar, Marienstr. 15}, \orgaddress{\state{Weimar}, \country{Germany}}}

\address[rvt6]{\orgdiv{Institute of Photonics, Department of Mathematics and Physics}, \orgname{Leibniz University}, \orgaddress{\state{ Hannover}, \country{Germany}}}

\address[rvt4]{\orgdiv{School of Mechanical, Medical and Process Engineering}, \orgname{Queensland University of Technology}, \orgaddress{\state{Brisbane}, \country{Australia}}}

\address[rvt7]{\orgdiv{Department of Applied Physics and Electronics}, \orgname{Umeå University}, \orgaddress{\state{Umea}, \country{Sweden}}}


\corres{Yinghua Liu, Department of Engineering Mechanics, Tsinghua University, Beijing 100084,
	China. \email{\url{yhliu@mail.tsinghua.edu.cn}}\\
	Xiang Li, School of Information Science and Technology, Hainan Normal University, Haikou, China. \email{\url{li\_xiang@hainnu.edu.cn}}}


\fundingInfo{Key Project of the National Natural Science Foundation of China (12332005)}

\abstract[Abstract]{Homogenization is a fundamental tool for studying multiscale physical phenomena. Traditional numerical homogenization methods, heavily reliant on finite element analysis, demand significant computational resources, especially for complex geometries, materials, and high-resolution problems. To address these challenges, we propose PreFine-Homo, a novel numerical homogenization framework comprising two phases: pretraining and fine-tuning. In the pretraining phase, a Fourier Neural Operator (FNO) is trained on large datasets to learn the mapping from input geometries and material properties to displacement fields. In the fine-tuning phase, the pretrained predictions serve as initial solutions for iterative algorithms, drastically reducing the number of iterations needed for convergence. The pretraining phase of PreFine-Homo delivers homogenization results up to 1000 times faster than conventional methods, while the fine-tuning phase further enhances accuracy. Moreover, the fine-tuning phase grants PreFine-Homo unlimited generalization capabilities, enabling continuous learning and improvement as data availability increases. We validate PreFine-Homo by predicting the effective elastic tensor for 3D periodic materials, specifically Triply Periodic Minimal Surfaces (TPMS). The results demonstrate that PreFine-Homo achieves high precision, exceptional efficiency, robust learning capabilities, and strong extrapolation ability, establishing it as a powerful tool for multiscale homogenization tasks.}




\maketitle



\section{Introduction} 

The homogenization method is a mathematical method for studying the macroscale performance of multiscale structures or materials \citep{mei2010homogenization}. Its core idea is to use a mathematical model to simplify the complex structural behavior on the microscale to a simple representation on the macroscale specifically an effective elastic tensor \citep{andreassen2014determine}. 
The homogenization method plays an important role in the field of mechanics and engineering \citep{charalambakis2010homogenization}. In mechanics, this approach allows researchers to predict the mechanical behavior at the macroscale based on the material structures at the microscale \citep{otero2015numerical}.
In this way, researchers can perform the mechanical analysis only on the macro-level model while maintaining an acceptance level of accuracy in the analysis,  significantly reducing the complexity and cost of the calculations \citep{hashin1962variational,hashin1963variational,hill1963elastic}. 
For example, researchers have applied homogenization methods to study the macroscopic material properties of various materials, such as fiber-reinforced composites \citep{tian2016numerical}, particulate composites \citep{okada2004homogenization}, laminated composites \citep{carvelli2001homogenization}; metamaterial \citep{zadpoor2016mechanical,pahlavani2024deep,saldivar2023bioinspired} such as metallic porous biomaterials \citep{bobbert2017additively}, photonic crystals \citep{halevi1999photonic}, phononic crystals \citep{antonakakis2013high}, auxetic materials  \citep{dos2012equivalent}, electromagnetic metamaterial \citep{ciattoni2015nonlocal}; porous media such as rock \citep{huang2011numerical}, wood \citep{diaz2019multiscale}, trabecular bone \citep{hollister1991application,makowski2013multiscale}, lattice materials \citep{weeger2021numerical,arabnejad2013mechanical}, various cellular materials \citep{duster2012numerical,nguyen2014computational,dong2019149}, functionally graded materials \citep{anthoine2010second,hasanov2021hierarchical}. As a result, the homogenization method is essential in science and engineering. 

The common branches of homogenization include direct, asymptotic, statistical, and numerical methods \citep{hill1963elastic,kalamkarov2009asymptotic,baniassadi2011mechanical,hassani1998review}. In the numerical (or computational) homogenization method, numerical methods such as the finite element method are usually used as a bridge between different scales for solving partial differential equations (PDEs). In the numerical homogenization method, numerical simulations and analyses are obtained by solving PDEs at the microscale level, and the numerical results are then converted into macroscopic material properties using homogenization formulas \citep{andreassen2014determine}. Various numerical tools for solving PDEs, including the finite element method \citep{andreassen2014determine}, the boundary element method \citep{okada2001homogenization}, the meshless method \citep{rodrigues2018homogenization}, discrete element method \citep{wellmann2008homogenization}, extended finite element method (XFEM) \citep{gal2013homogenization}, spectral method \citep{bellis2024numerical}, and molecular dynamics \citep{shin2013multiscale} have been incorporated with numerical homogenization. However, the numerical computation of PDEs, especially using the finite element method, plays the most important role in the process of numerical homogenization, yet it involves a significantly large amount of computation. Therefore, reducing the computational load during numerical homogenization is of critical importance.

The recent advancements in the field of AI for PDEs have introduced a new data-driven paradigm, significantly reducing the computational time for solving PDEs \citep{hao2020ai}. Given that solving PDEs is a critical step in homogenization, employing AI technologies has the potential to greatly increase the computational efficiency of homogenization, which serves as the motivation for our proposed model of homogenization.
We introduce some of the latest research in AI for PDEs \citep{wang2024artificial,yizheng2024ai}.
A groundbreaking approach introduced in 2019 is based on Physics-Informed Neural Networks (PINNs) \citep{PINN_original_paper}, presenting a novel scientific computing method using deep learning \citep{cuomo2022scientific}. PINNs integrates data and physical equations through neural networks and exhibits unique advantages in inverse problems and high-dimensional scenarios \citep{PINN_review}.
Recently, operator learning has gained considerable attention \citep{DeepOnet}. Notable works in neural operators include DeepONet \citep{DeepOnet} and Fourier neural operator (FNO) \citep{li2020fourier}. Initially, operator learning was essentially a data-driven approach, differing from traditional methods by possessing reliable mathematical support and discrete invariance \citep{kovachki2023neural}. The theoretical foundation of DeepONet is based on a universal approximation to nonlinear operators \citep{chen1995universal} that neural networks can approximate any continuous operator. FNO, on the other hand, is grounded in Fourier transforms \citep{de2022generic,kovachki2021universal}. Discrete invariance implies the ability to train and test on arbitrary grids without the need for retraining when the grid discretization changes. This feature is important in practical engineering applications \citep{wen2022u}, especially in our numerical homogenization process.
The varying resolution of our data benefits from the discrete invariance property, allowing the model to learn from data of different resolutions.
The most significant advantage of employing operator learning to solve PDEs is the substantial acceleration of computations, and the efficiency of operator learning is several orders of magnitude higher than traditional numerical algorithms \citep{hao2020ai}, especially in numerical weather prediction  \citep{pathak2022fourcastnet,bi2023accurate}.
The recent advancements in operator learning involve integrating physics equations with operator learning, known as Physics-Informed Neural Operator (PINO) \citep{goswami2022physics,li2024physics,wang2021learning,wang2023dcm}. PINO utilizes operator learning to achieve an initial solution that closely approximates the true solution, and then leverages physical equations for fine-tuning from this initial solution to attain a solution with enhanced accuracy.

Therefore, we are motivated to integrate operator learning into the homogenization process to reduce the computational load by several orders of magnitude.
In this study, we have chosen to illustrate the performance of the proposed numerical homogenization method using a type of cellular material. Cellular materials are characterized by porous microstructures leading to low weight and exceptional mechanical performance. The superior specific material properties and engineering values of natural cellular materials such as wood and cork have motivated the development of artificial cellular materials to achieve high specific stiffness \citep{liu2024ultrastiff} and \citep{schaedler2011ultralight,zheng2014ultralight}, strength \citep{bauer2014high}, negative Poisson’s ratio \citep{zhang2016mechanical,grima2006negative}, energy absorption \citep{tay2014finite}\citep{xu2018review}, thermal insulation \citep{li2021biomimetic}, and acoustic manipulation \citep{al2013vibro}. Advances in artificial cellular materials are paving the way for applications in aerospace \citep{hohe2012numerical}, automotive \citep{baroutaji2019application}, medical implants \citep{benedetti2021architected,baroutaji2021metallic}, soft robotics \citep{lee20203d}\citep{goswami20193d} and electronic devices \citep{jiang2018auxetic}. 
Triply Periodic Minimal Surfaces (TPMS) is a type of artificial cellular material with topology-driven properties. The concept of TPMS was first proposed by Hermann Schwarz in 1865 \citep{schwarz1972gesammelte}. Several commonly used TPMS models are Schwarz Primitive, Schwarz Diamond, Neovious, Schoen Gyroid, Schoen IWP, Schoen FRD, Fischer Kosh S \citep{al2019multifunctional}. TPMS surfaces are spatially periodic in all three dimensions and have zero mean curvature \citep{han2018overview}. TPMS materials are characterized by exceptional strength-to-weight and surface-to-volume ratios and have therefore been successfully used in various engineering applications, including artificial implants \citep{barba2019synthetic}, heat exchangers \citep{attarzadeh2022multi}, energy absorption \citep{yu2019investigation}, soft robotics \citep{hussain2020design}, energy storage \citep{qureshi2021heat}, sound absorption \citep{zhang2023sound}. The homogenization method has been successfully used to characterize the effective elastic tensor of TPMS \citep{modrek2022optimization, fu2022isotropic, pais2023multiscale}.
Therefore, it is important to integrate modern neural operators into the numerical homogenization method to accelerate the process. 
Huang et al. \citep{huang2023introduction} proposed a framework using neural operators for multiscale homogenization, but this framework relies on data and is limited to 2D problems.

To address these limitations, \textbf{we propose a novel numerical homogenization framework that integrates data-driven approaches with physical equations for 3D elastic problems}, named PreFine-Homo. 
In particular, our framework exhibits three major characteristics:
\begin{enumerate}
	\item Our framework consists of two phases: pretraining and fine-tuning. The pretraining phase leverages large datasets to provide a suitable initial solution, while the fine-tuning phase incorporates physical equations for refinement. By combining data and physical equations in these two phases, PreFine-Homo forms a self-learning framework. Importantly, PreFine-Homo can operate even without data, relying solely on physical equations for homogenization.
	
	\item Our framework is highly flexible, with the fine-tuning phase being optional. If high precision is not required, the fine-tuning phase can be skipped. However, if high precision is desired, the prediction from the pretraining phase can be used as the initial solution for the iterative algorithm in the fine-tuning phase, significantly reducing the number of iterations. The fine-tuning phase trades additional computational time for higher accuracy.
	
	\item The pretraining phase of our framework demonstrates high accuracy and exceptional efficiency, being 1,000 times faster in predicting the displacement field necessary for homogenization compared to the Finite Element Method (FEM). The fine-tuning phase of our framework provides extrapolation capabilities, enabling the calculation of effective elastic tensors even for data that significantly deviates from the pretraining dataset.
	
	\item Our framework is capable of performing rapid homogenization calculations for any geometric shape, Poisson's ratio, and resolution in Triply Periodic Minimal Surfaces (TPMS). The framework can be trained on low-resolution datasets and tested on high-resolution datasets, showcasing its flexible learning capability.
\end{enumerate}

The core of our framework involves two steps: the pretraining phase and the fine-tuning phase. In the pretraining phase, we use traditional PDE solvers to obtain large datasets. Then, we employ the Fourier Neural Operator (FNO) to fit the mapping from geometric shapes and Poisson's ratios to displacement fields. 
In the optional fine-tuning phase, we use the prediction from the pretraining phase as the initial solution for the iterative algorithm. Since the pretraining phase is based on large datasets, the initial solution provided by pretraining is close to the true solution, thereby significantly reducing the number of iterations and improving computational efficiency.
Finally, based on the obtained displacement field, we apply the homogenization formula to derive the final effective elastic tensor. The pretraining and fine-tuning phases of PreFine-Homo are conceptually similar to the pretraining and reinforcement learning stages of large language models \citep{guo2025deepseek}.

The outline of the paper is as follows. \Cref{sec:tradition_numerical_homo} introduces the steps of traditional homogenization. \Cref{sec:method} describes the datasets obtained from finite element calculations, the FNO algorithm, and our proposed PreFine-Homo model. \Cref{sec:Result} provides a comprehensive validation of our proposed PreFine-Homo model across different geometric shapes, various Poisson's ratios, different data resolutions, and different distributions of training and test sets. \Cref{sec:Discussion} presents some discussion.  Finally, \Cref{sec:Conclusion} summarizes the advantages and limitations of our proposed homogenization framework PreFine-Homo, and offers an outlook on future research directions.

\section{Preparatory knowledge} \label{sec:tradition_numerical_homo}

In this section, we introduce the process of traditional numerical homogenization. Andreassen et al. \cite{andreassen2014determine} proposed a traditional numerical homogenization method to calculate the effective macroscopic elastic tensor of periodic composite material. 
We will briefly introduce this method below.
For simplicity, we consider a material consisting of solid material and void. \( E_{ijkl}^{S} \) denotes the isotropic elastic matrix of the solid material. For calculating the homogenized elastic matrix \( E_{ijkl}^{H} \), we applied 6 unit strain fields over the unit cell, including three normal strain and three shear strain fields. The 6 unit strain fields $\varepsilon_{i}^{macro}$ are defined as
\begin{equation}
	\begin{aligned}
		\varepsilon_{i}^{\text{macro}} = \begin{pmatrix} \delta_{1i} & \delta_{2i} & \delta_{3i} & \delta_{4i} & \delta_{5i} & \delta_{6i} \end{pmatrix}^T \quad \\
		\text{for } i = 1,2,3,4,5,6
		\label{eq:unit strain}
	\end{aligned}
\end{equation}
From 1 to 6, they correspond to normal strains in the x, y, and z directions, and shear strains in the xy, xz, and yz directions, respectively.
$\delta_{1i}$ is Kronecker delta (I=1,2,3,4,5,6). For instance,
\begin{equation}
	\varepsilon_{1}^{\text{macro}} = \begin{pmatrix} 1 & 0 & 0 & 0 & 0 & 0 \end{pmatrix}^T \quad \text{for } i = 1
\end{equation}
The homogenized elastic matrix \( E_{ijkl}^{H} \) is calculated according to the following equation 
\begin{equation}
	\begin{aligned}
		E_{ijkl}^{H} = \frac{1}{\vert \Omega \vert} \int_{\Omega} E_{pqrs} \left( \varepsilon_{pq}^{(0)(ij)} - \varepsilon_{pq}^{(ij)} \right)\\
		*\left( \varepsilon_{rs}^{(0)(kl)} - \varepsilon_{rs}^{(kl)} \right) d\Omega
		\label{eq:homogenized elastic matrix}    
	\end{aligned}
\end{equation}
In this equation, $E_{pqrs}$ denotes the local elastic matrix of the TPMS unit cell, and $E_{pqrs}$ obeys
\begin{equation}
	E_{ijkl} = 
	\begin{cases} 
		{E}_{ijkl}^{S} & \text{for solid material}, \\
		0 & \text{for void}.
	\end{cases}
\end{equation}
$\vert\Omega\vert$ represents the volume of the unit cell domain.  $\varepsilon_{pq}^{(0)(ij)}$ shown in \Cref{eq:unit strain}, denotes a macroscopic unit strain field imposed on $\Omega$. 
$\varepsilon_{pq}^{(ij)}$, on the other hand, denotes the local strain field calculated by
\begin{equation}
	\epsilon^{(ij)}_{pq} = \epsilon_{pq} \left( \boldsymbol{X}^{ij} \right) = \frac{1}{2} \left( X^{ij}_{p,q} + X^{ij}_{q,p} \right)
\end{equation}
In this equation, $\boldsymbol{X}^{ij}$ denotes the real displacement field under by applying the macroscopic unit strain field over  $\Omega$. $\varepsilon_{pq} (\boldsymbol{X}^{kl})$ is calculated by solving the following elasticity equation
\begin{equation}
	\begin{aligned}
		\int_{\Omega} E_{ijkl} \varepsilon_{ij} (\boldsymbol{X}^{*}) \varepsilon_{pq} (\boldsymbol{X}^{kl}) d\Omega = \\
		\int_{\Omega} E_{ijkl} \varepsilon_{ij} (\boldsymbol{X}^{*}) \varepsilon^{0(kl)}_{pq} d\Omega    
	\end{aligned}
\end{equation}
, where  $\boldsymbol{X}^{*}$ refers to a virtual displacement field. The finite element method is usually used to solve the real displacement field  $\boldsymbol{X}^{kl}$ in the above equation. Then, the actual strain $\boldsymbol{\varepsilon}^{(ij)}$ can be calculated from $\boldsymbol{X}^{ij}$ and finally substituted into \Cref{eq:homogenized elastic matrix}  to solve the homogenized elastic matrix $E_{ijkl}^{H}$.

We observe that the process of homogenization necessitates the computation of displacement fields under six different load conditions. Traditional approaches rely on finite element analysis introduced above, consuming substantial amounts of time for computation. Different geometry and materials should be recalculated to get the corresponding displacement field. Thus, enhancing the speed of calculating displacement fields naturally leads to a significant improvement in the efficiency of the total homogenization process. Next, we explain how we use operator learning to accelerate homogenization calculations. 

\section{Methods} \label{sec:method}
Our discussion is structured into three main sections. Initially, we introduce the dataset calculated from traditional homogenization methods using finite element method (FEM). Subsequently, we present the Fourier neural operator.
Lastly, we elaborate on the integration of FNO with the homogenization process, i.e., PreFine-Homo.

\subsection{Dataset}\label{sec:dataset}
Our data, computed with a supercomputer (900 CPU nodes each with 192G of memory and 80 V100 GPUs), includes 1800 sets of TPMS geometries each with 2,097,152 elements (resolution 128*128*128) and 3600 sets each with 262,144 elements (resolution 64*64*64), covering various geometries and Poisson's ratios, totaling approximately 325G of data. Computing data at a resolution of 128 on a system with 192 GB of memory and 56 CPU cores requires an average of 2500 seconds (total CPU time of 128: 1800*2500=52 days). Due to the iterative method of our custom-written traditional finite element homogenization code, the time for traditional numerical homogenization is not fixed, so we can only provide an average time. For data at a resolution of 64, under the same hardware conditions, the average computation time is 150 seconds (total CPU time of 64: 3600*150=6.25 days). 

Notably, the computation times mentioned above are based on the Finite Element Method (FEM). Additionally, we did not use finite element analysis software like Abaqus to obtain data; instead, we developed the numerical homogenization code in MATLAB and Python.

Furthermore, while solving periodic problems using the Fast Fourier Transform (FFT) is theoretically faster than using FEM, the efficiency of FFT-based is less stable compared to FEM. Given that FEM is more versatile and widely applicable, we chose to use FEM for data generation and as the benchmark for comparison with our proposed PreFine-Homo framework. Detailed results about FFT to solve periodic problems are provided in \Cref{sec:FFT}.
\subsection{Fourier neural network for homogenization}

FNO has garnered significant attention as an algorithm in neural operator learning. Therefore, before delving into the details of FNO, let's first introduce neural operator learning \citep{kovachki2023neural} for a better understanding of FNO.

Neural operator learning is a data-driven algorithm widely employed in science and engineering, particularly for mapping relationships between different functions. The advantage of the neural operator is the invariance to discretization, and  the concept of discretization-invariant include:

\begin{enumerate}
	\item The ability to handle input data of arbitrary resolutions.
	\item The capability to produce output at any given location.
	\item Convergence of results as the grid is refined.
\end{enumerate}

In contrast to conventional data-driven methods based on neural networks, which are often sensitive to discretization levels and may require retraining when changing input-output function resolutions, neural operator learning exhibits invariance to discretization. This property makes it a valuable data-driven algorithm model, as highlighted by its practical applications \citep{wen2022u}.
The feature of invariance to discretization means that we can train the model in the coarse mesh and predict the result in the fine mesh. 
It is a great advantage in practice that training data can possess arbitrary resolution.

Moreover, not all fitting algorithms can be classified as neural operators. Neural operators not only need to satisfy the invariance to discretization but also meet the requirements of generalized approximation for continuous operators \citep{chen1995universal}. 
This characteristic ensures the advantage of neural operators in learning the family of PDEs. Through operator learning, we can grasp the implicit mappings of PDEs, meaning the rapid prediction of solution spaces from input spaces (such as boundary conditions, geometric shapes, and material properties). We can conceptualize PDEs as implicit mappings from input spaces to solution spaces. Because this mapping is challenging to represent explicitly, the traditional PDEs solver can be used to obtain the data for approximating the implicit mapping. The tool for approximating the implicit mapping can be a neural operator due to its powerful approximation capabilities.

Among neural operators including Graph Neural Operator (GNO) \citep{li2020neural}, Local Neural Operator (LNO) \citep{kovachki2023neural}, and Fourier Neural Operator (FNO) \citep{li2020fourier}, the Fourier neural operator (FNO) in \Cref{fig:FNO_intro} stands out as a category with superior performance. The algorithmic design of neural operator learning revolves around the core operation of kernel integration within the neural operator layer:

\begin{figure}
	\begin{centering}
		\includegraphics[scale=0.60]{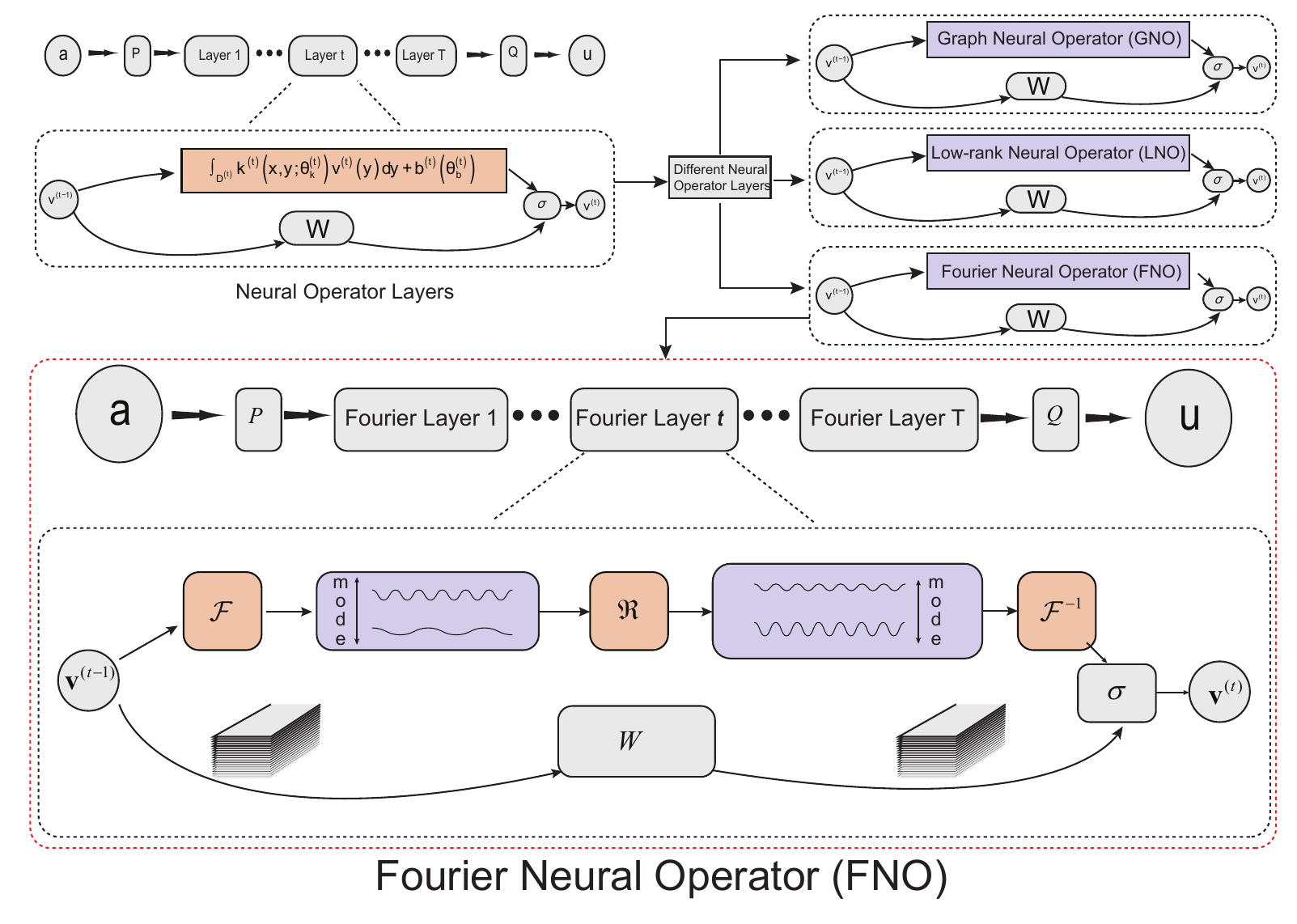}
		\par\end{centering}
	\caption{Schematic diagram of the Fourier neural operator (FNO) in the neural operator including GNO, LNO, and FNO: $\boldsymbol{P}$ is used to upsample the input data for better processing by the Fourier layer. $\boldsymbol{Q}$ downsamples the data processed by the Fourier layer to the output space. $\boldsymbol{v}^{(t-1)}$ and $\boldsymbol{v}^{(t)}$ are the input and output of the $t$-th Fourier layer, respectively. $\mathcal{F}$ is the fast Fourier transform. $\Re$ performs a linear transformation on the output after the fast Fourier transform, and then applies the inverse Fourier transform $\mathcal{F}^{-1}$. $\boldsymbol{W}$ performs a linear transformation on $\boldsymbol{v}^{(t-1)}$. $\sigma$ is a nonlinear transformation.
		\label{fig:FNO_intro}}
\end{figure}

\begin{equation}
	\mathcal{\boldsymbol{H}}^{(t)}(\boldsymbol{v}^{(t)};\boldsymbol{\theta}_{k}^{(t)})(\boldsymbol{x})=\int_{D^{(t)}}\boldsymbol{k}^{(t)}(\boldsymbol{x},\boldsymbol{y};\boldsymbol{\theta}_{k}^{(t)})\boldsymbol{v}^{(t)}(\boldsymbol{y})d\boldsymbol{y},\label{eq:kernel function}
\end{equation}
where $\boldsymbol{v}^{(t)}$ represents the input to the neural operator layer, $\boldsymbol{k}^{(t)}(\boldsymbol{x},\boldsymbol{y}; \boldsymbol{\theta}_{k}^{(t)})$ is the kernel function approximated using a neural network with parameters $\boldsymbol{\theta}_{k}^{(t)}$. The kernel integration algorithm transforms  the function $\boldsymbol{v}^{(t)}$ through integration into another type of function $\mathcal{\boldsymbol{H}}^{t}(\boldsymbol{v}^{(t)})(\boldsymbol{x})$ by the action of the kernel operator $\mathcal{\boldsymbol{H}}^{t}$.  The index $t$ refers to the $t$-th layer of the neural operator. Consequently, the overall computation flow of operator learning is expressed as:

\begin{equation}
	\begin{alignedat}{1}\boldsymbol{G}_{\theta}(\boldsymbol{a}(\boldsymbol{x});\boldsymbol{\theta})= & \boldsymbol{Q}(\boldsymbol{v}^{(T+1)};\boldsymbol{\theta}_{Q})\circ\sigma[\mathcal{\boldsymbol{H}}^{(T)}(\boldsymbol{v}^{(T)};\boldsymbol{\theta}_{k}^{(T)})\\
		& +\boldsymbol{W}^{(T)}(\boldsymbol{v}^{(T)};\boldsymbol{\theta}_{w}^{(T)})+\boldsymbol{b}^{(T)}(\boldsymbol{\theta}_{b}^{(T)})]\circ\\
		& \cdots\circ\sigma[\mathcal{\boldsymbol{H}}^{(1)}(\boldsymbol{v}^{(1)};\boldsymbol{\theta}_{k}^{(1)})+\\
		& \boldsymbol{W}^{(1)}(\boldsymbol{v}^{(1)};\boldsymbol{\theta}_{w}^{(1)})+\boldsymbol{b}^{(1)}(\boldsymbol{\theta}_{b}^{(1)})]\circ\\
		& \boldsymbol{P}(\boldsymbol{a}(\boldsymbol{x});\boldsymbol{\theta}_{P})
		\label{eq:neural operator algorithm}
	\end{alignedat},
\end{equation}
where $\boldsymbol{P}$ and $\boldsymbol{Q}$ are responsible for dimensionality expansion and reduction of the original data, and their respective learnable parameters are $\boldsymbol{\theta}_{P}$ and $\boldsymbol{\theta}_{Q}$. $\mathcal{\boldsymbol{H}}^{(t)}$, $\boldsymbol{W}^{(t)}$, and $\boldsymbol{b}^{(t)}$ represent kernel integration, linear transformation, and bias, with corresponding learnable parameters $\boldsymbol{\theta}_{k}^{(t)}$, $\boldsymbol{\theta}_{w}^{(t)}$, and $\boldsymbol{\theta}_{b}^{(t)}$. It is noteworthy that these learnable parameters are typically approximated using neural networks. 

Different kernel integration algorithms give rise to various operator learning algorithms, making the design of kernel integration algorithms a core aspect of operator learning. FNO is the specific instantiation of kernel integration algorithms using Fourier transforms. Theorems suggest that the formulation in \Cref{eq:neural operator algorithm} meets the requirements of generalized approximation \citep{de2022generic,kovachki2021universal}, indicating that operator learning algorithms designed according to  \Cref{eq:neural operator algorithm} possess favorable convergence properties.

Fourier Neural Operator (FNO) is a specific neural operator \citep{li2020fourier} that has garnered significant attention due to its efficiency and accuracy. Its core idea involves integrating Fourier transform into the neural operator, specifically representing  \Cref{eq:kernel function} using Fourier transform:

\begin{equation}
	\begin{aligned}
		\mathcal{\boldsymbol{H}}^{(t)}(\boldsymbol{v}^{(t)};\boldsymbol{\theta}_{k}^{(t)})(\boldsymbol{x})&=\int_{D^{(t)}}\boldsymbol{k}^{(t)}(\boldsymbol{x},\boldsymbol{y};\boldsymbol{\theta}_{k}^{(t)})\boldsymbol{v}^{(t)}(\boldsymbol{y})d\boldsymbol{y}\\
		& =\mathcal{F}^{-1}\circ\Re\circ\mathcal{F}(\boldsymbol{v}^{(t)}(x)),\label{eq:kernel function-1}   
	\end{aligned}
\end{equation}
where $\mathcal{F}$ and $\mathcal{F}^{-1}$ are the Fourier transform and its inverse, respectively, and $\Re$ denotes linear transformation.
To explain the FNO algorithm's process in detail, we combine the FNO algorithm framework with a specific example.  Considering a two-dimensional steady-state heat conduction equation:

\begin{equation}
	\begin{cases}
		\text{Domain:} & -\nabla\cdot(K(x, y)\nabla T(x,y)) =f(x,y)\\
		\text{Boundary:} & T(x,y)=u(x,y) \\
	\end{cases},\label{eq:burgers equations}
\end{equation}
where $T(x,y)$ is the temperature field to be solved, $K(x,y)$ is the non-uniform thermal conductivity field, and once the boundary condition $u(x,y)$, thermal conductivity $K(x,y)$, and heat source $f(x,y)$ are determined, the solution $T(x,y)$ can be uniquely determined. In this problem, the input is the thermal conductivity $K\in\mathbb{R}^{d_{i}}$, where $d_{i}$ is the total sample points of thermal conductivity, e.g., $d_{i}=128*128$ If there are evenly 128 points on each side of the two-dimensional heat conduction problem.  In this example, the thermal conductivity $K$ is equivalent to $a(x)$ in \Cref{eq:neural operator algorithm} in FNO, and the output is the steady-state temperature field $T(x,y)$ if the boundary condition is fixed. The FNO algorithm proceeds by first using a fully connected neural network $P$ to lift the input $a(x)$ to $d_{c}$ dimensions (i.e., $P:\mathbb{R}\rightarrow\mathbb{R}^{d_{c}}$, where $d_{c}$ is analogous to channels in computer vision):

\begin{equation}
	\boldsymbol{v}^{(0)}(x)=P(a(x))\in\mathbb{R}^{d_{i}*d_{c}}\text{.}
\end{equation}

Next, $\boldsymbol{v}^{(0)}(x)$ is passed through Fourier Layer 1. Firstly, a Fourier transform is applied to each channel of $\boldsymbol{v}^{(0)}(x)$, resulting in

\begin{equation}
	\boldsymbol{f}^{(1)}(x)=\mathcal{F}(\boldsymbol{v}^{(0)}(x))\in\mathbb{\mathbb{C}}^{d_{i}*d_{c}}\text{.}
\end{equation}

High-frequency components are filtered out, retaining only low frequencies by truncating at a frequency of $d_{m}$ ($1\leq d_{m}\leq d_{i}$). Therefore, the data structure transforms to $\boldsymbol{f}^{(1)}(x)\in\mathbb{\mathbb{C}}^{d_{m}*d_{c}}$. A linear transformation $\Re$ ($\Re:\mathbb{R}^{d_{c}}\rightarrow\mathbb{R}^{d_{c}}$) is then applied to each frequency channel:

\begin{equation}
	\boldsymbol{f}_{L}^{(1)}(x)=\Re(\boldsymbol{f}^{(1)}(x))\in\mathbb{\mathbb{C}}^{d_{m}*d_{c}}, \label{eq:linear_transformation}
\end{equation}
where there are $d_{m}$ different linear transformation matrices. Because different frequencies use different linear transformations, a set of linear transformation matrices $\mathbb{R}^{d_{c}*d_{c}*d_{m}}$ is employed. After the linear transformation $\Re$, the dimension structure of the data remains unchanged. Since Fourier inverse transform is required in the next step, zero-padding is performed to reintroduce the high-frequency components removed during truncation.  An inverse Fourier transform to $\boldsymbol{f}_{L}^{(1)}(x)$ is applied:

\begin{equation}
	\boldsymbol{v}_{f}^{(1)}=\mathcal{F}^{-1}(\boldsymbol{f}_{L}^{(1)}(x))\in\mathbb{\mathbb{\mathbb{R}}}^{d_{i}*d_{c}}\text{.}
\end{equation}

The advantage of the $\Re$ linear transformation lies in learning the mapping relationships between data in the Fourier frequency space. Furthermore,  $\Re$ acts as a form of regularization due to high-frequency filtering, reducing overfitting and enhancing the model's generalization.

Additionally, following the idea of residual networks \citep{he2016deep}, a linear transformation is directly applied to the channels of $\boldsymbol{v}^{(0)}(x)$. This linear transformation, which is the same for points, involves only one linear transformation $W\in\mathbb{R}^{d_{c}*d_{c}}$:

\begin{equation}
	\boldsymbol{v}_{L}^{(1)}=W(\boldsymbol{v}^{(0)}(x))\in\mathbb{\mathbb{\mathbb{R}}}^{d_{i}*d_{c}}\text{,} \label{eq:W}
\end{equation}
$W$ uses the residual network concept to reintroduce high-frequency components removed by the Fourier transform into the network for learning. As a result, the residual connection of $W$ prevents a degradation in predictive performance caused by the removal of high-frequency components.

The results of Fourier transform and linear transformation are combined:

\begin{equation}
	\boldsymbol{v}^{(1)}(x)=\sigma(\boldsymbol{v}_{f}^{(1)}+\boldsymbol{v}_{L}^{(1)}+\boldsymbol{b}^{(1)})\in\mathbb{\mathbb{\mathbb{R}}}^{d_{i}*d_{c}},
\end{equation}
where $\boldsymbol{b}^{(1)}$ is the bias. The introduction of non-linear transformations \(\sigma\) not only increases the expressive power of the network but also simultaneously enhances the fitting capability of high frequencies. The Fourier layer is the core of FNO's computation. The above operations are repeated T times to obtain $\boldsymbol{v}^{(T)}(x)$, which is then reduced to the target dimension by $Q$. In the case of the steady-state heat conduction problem, $Q:\mathbb{R}^{d_{c}}\rightarrow\mathbb{R}$, and FNO computation is over.

FNO has two major advantages: firstly, due to Fourier transformation filtering out high-frequency modes, it can improve the model's speed and generalization ability, reducing overfitting; secondly, it exhibits discretization-invariant. Since all FNO operations are point-wise computations independent of the data grid, it possesses the advantage of discretization-invariant.

FNO layers are repeated \(T\) times, followed by dimension reduction to obtain the final output. 
Below, we illustrate how to predict the effective elastic tensor in the proposed PreFine-Homo with FNO.

\subsection{PreFine-Homo: Pretraining phase}
The crucial steps in calculating the effective elastic tensor involve obtaining the displacement field $\boldsymbol{u}$ in six load cases. However, traditional methods involve obtaining the displacement field through the finite element method (FEM) six times due to six load cases. For different geometric or material configurations (such as Poisson's ratio or Young's modulus), we need to recompute using FEM. As the material becomes more heterogeneous and complex, the computational demands of FEM become a bottleneck in efficiency. Therefore, enhancing the speed of displacement prediction is essential for improving the efficiency of calculating the effective elastic tensor in homogenization.

\begin{figure}
	\begin{centering}
		\includegraphics[scale=0.40]{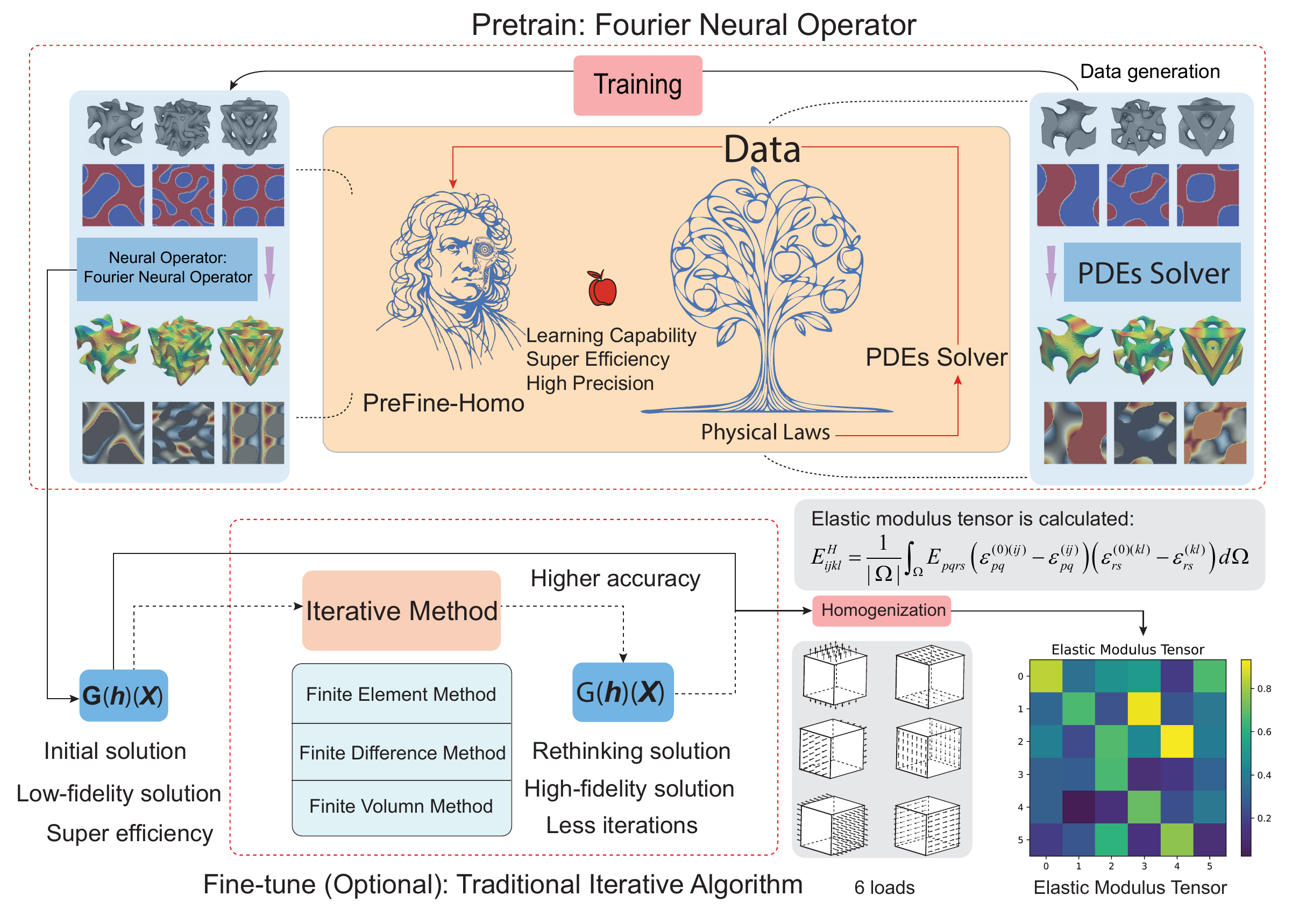}
		\par\end{centering}
	\caption{PreFine-Homo consists of two key steps: \textbf{pretraining} and \textbf{fine-tuning}. In the pretraining phase, a large amount of data with different materials and geometries is obtained using traditional numerical solvers. Subsequently, the Fourier Neural Operator (FNO) is employed to fit the large dataset. The fine-tuning phase involves using the FNO to provide an initial solution, which is then used as the initial vector in a traditional iterative algorithm for further refinement. 
		The pretraining phase of PreFine-Homo enables rapid prediction of displacement fields for new geometries and materials in the test set. Depending on the requirements, the fine-tuning phase may be optionally applied, and the effective elastic tensor is subsequently obtained based on the homogenization formula.
		\label{fig:method}}
\end{figure}

Here, we adopt the Fourier neural operator (FNO) \citep{li2020fourier} as a backbone in our proposed PreFine-Homo, as shown in the pretrain phase of \Cref{fig:method}, because FNO is an algorithm that has garnered significant attention in AI for PDEs, especially in weather prediction \citep{bi2023accurate,pathak2022fourcastnet}. Given the success of FNO in fluid mechanics weather prediction, it is reasonable to believe that FNO can also achieve success in solid mechanics, as both fundamentally involve function mapping theoretically. 

The core concept of our approach is straightforward: we first obtain a big dataset with various Poisson's ratios and geometric configurations as input data, and the 18 displacement fields subjected to six different loading conditions as output data (three directions: x, y, and z for each of the six types of loading). Training is conducted using the Fourier neural operator (FNO), with a material matrix as the input and 18 displacement fields as the output. The material matrix is a 3D matrix of resolution 128*128*128 and 64*64*64, populated with zeros and Poisson's ratios. Given our focus on linear elasticity homogenization, we set Young's modulus to 1 during training, simplifying the final calculation of the effective elastic tensor by linearly multiplying by Young's modulus.

Using the dataset obtained from the traditional homogenization process, we began training our PreFine-Homo. The training time for a resolution of 128 is about 5 hours per 100 epochs (3D FNO with mode 18 on one V100: 32G), and the training time for a resolution of 64 is about 3 hours per 100 epochs (3D FNO with mode 32 on one V100: 32G). The mode of FNO and the amount of data can be configured according to hardware conditions. The model takes a material matrix with four channels as input—specifically, one for the material's Poisson's ratio and three for the spatial coordinates (x, y, and z)—and produces an 18-channel displacement field as output. To address the varying magnitudes across different displacement fields, we implemented normalization techniques to ensure consistency across the displacement channels. Our model is capable of rapidly predicting the 18 necessary displacement fields for homogenization when given any Poisson's ratio within the range of 0.1 to 0.4, any value of elastic modulus, and any geometry of the TPMS (Triply Periodic Minimal Surface) structure. It is important to note that we focused on Poisson's ratios ranging from 0.1 to 0.4 because this range is most commonly encountered in engineering applications.

After obtaining the displacement fields required for homogenization, we incorporate them into the conventional homogenization process in \Cref{eq:homogenized elastic matrix}, to ultimately derive the effective elastic tensor.

Given PreFine-Homo's capability to assimilate data of any resolution, the model has remarkable learning capabilities, which means the model can continually incorporate data, ultimately evolving into a supermodel capable of homogenizing across any geometry and material. The model excels in high precision and super efficiency, particularly in terms of efficiency.

\subsection{PreFine-Homo: Fine-tuning phase}

During the pretraining phase, PreFine-Homo is trained on a large dataset, which often results in highly efficient performance during the testing phase. However, since the training is always based on high-precision data, the theoretical accuracy of the FNO's predictions can only asymptotically approach that of the high-precision data. This implies that there will inevitably be some unavoidable errors. These errors primarily stem from two sources: (1) the data errors in the high-precision data itself, and (2) the model errors introduced by the FNO model when fitting the data.

To address this, we introduce the fine-tuning phase in PreFine-Homo, as shown in the fine-tuning phase of \Cref{fig:method}. The core idea of the fine-tuning phase is to use the FNO's prediction as the initial value for an iterative algorithm. Since the operator learned during the pretraining phase has been exposed to a vast amount of data, the model can provide a highly accurate initial solution. This initial solution is much closer to the true value compared to a random prediction, thereby significantly reducing the number of iterations required. Note that, theoretically, the fine-tuned solution can achieve arbitrary precision, provided that the threshold in the iterative algorithm is set sufficiently small.

The iterative algorithm used in the fine-tuning phase is determined based on traditional numerical methods for solving PDEs. In computational mechanics, these traditional numerical methods primarily include the Finite Element Method (FEM), Finite Difference Method (FDM), and Finite Volume Method (FVM). The choice of traditional numerical method determines the matrix equation in the iterative algorithm:
\begin{equation}
	\boldsymbol{K}\boldsymbol{X} = \boldsymbol{f}, \label{eq:linear_KXB}
\end{equation}
where $\boldsymbol{K}$ is determined by the selected numerical method, meaning that different numerical methods for solving PDEs will result in different $\boldsymbol{K}$. Here, $\boldsymbol{X}$ is the vector to be solved, and $\boldsymbol{f}$ is the known vector.

There are two numerical approaches to solving \Cref{eq:linear_KXB}: direct methods and iterative methods. For large-scale problems, iterative methods are often preferred due to their computational complexity. The computational complexity of direct and iterative methods is discussed in \Cref{sec:computational_time_iter}. Since we are considering 3D problems with over one million degrees of freedom, iterative methods are more suitable than direct methods due to their lower complexity. Various iterative algorithms can be employed, and some commonly used ones are listed in \Cref{sec:iterative_method}. Since the 3D elastic problem involves large, sparse systems of linear equations, our proposed PreFine-Homo framework employs the Preconditioned Conjugate Gradient (PCG) method detailed in \Cref{sec:iterative_method} as its iterative solver. The iterative formulation for solving \Cref{eq:linear_KXB} is:
\begin{equation}
	\boldsymbol{X}^{(k+1)} = \boldsymbol{\phi}(\boldsymbol{X}^{(k)}; \boldsymbol{K}, \boldsymbol{f}), \label{eq:iterative_formula}
\end{equation}
where $\boldsymbol{X}^{(k)}$ is the solution vector at the $k$-th iteration. The iteration stops when the relative error $\mathcal{L}_{rel}$ falls below a threshold $tol$:
\begin{equation}
	\mathcal{L}_{rel} = \frac{\|\boldsymbol{K}\boldsymbol{X} - \boldsymbol{f}\|}{\|\boldsymbol{f}\|} < tol, \label{eq:tol}
\end{equation}
where $\|\cdot\|$ denotes the $L_2$-norm.

The core of the fine-tuning phase in PreFine-Homo is to use the pretrained output $\boldsymbol{X}^{fno}$ as the initial solution for \Cref{eq:iterative_formula}:
\begin{equation}
	\boldsymbol{X}^{(k+1)} = \boldsymbol{\phi}(\boldsymbol{X}^{(k)}; \boldsymbol{K}, \boldsymbol{f}, \boldsymbol{X}^{(0)} = \boldsymbol{X}^{fno}).
\end{equation}

The pretraining and fine-tuning phases of PreFine-Homo are conceptually similar to the pretraining and reinforcement learning stages of large language models \citep{guo2025deepseek}. Since the pretraining phase of PreFine-Homo provides a high-quality initial solution, the number of iterations in the fine-tuning phase can be significantly reduced compared to traditional methods that use random initial vectors. A reduction in the number of iterations means a decrease in computational time. Theoretically, the fine-tuned solution can achieve arbitrary precision, provided that the convergence threshold in the iterative algorithm is set sufficiently small. It is worth noting that the fine-tuning phase is optional and can be applied based on specific requirements.
To determine whether the fine-tuning phase is necessary, we propose evaluating the initial solution provided by the pretraining phase of PreFine-Homo by substituting it into \Cref{eq:tol} and examining the value of  
$\mathcal{L}_{rel} = \|\boldsymbol{K}\boldsymbol{X}^{fno} - \boldsymbol{f}\|/\|\boldsymbol{f}\|$.
Since $\mathcal{L}_{rel}$ reflects the accuracy of the initial solution, if $\mathcal{L}_{rel} < tol_{fine}$, the fine-tuning phase is deemed unnecessary, where $tol_{fine}$ serves as the threshold for deciding whether to proceed with the fine-tuning phase.

If a certain level of error is acceptable, the pretrained model of PreFine-Homo can be used directly. However, if high-precision solutions are desired, the fine-tuning phase can be employed at the cost of additional computational resources.

PreFine-Homo combines data-driven and physics-based approaches, embodying a dual-thinking paradigm. The framework of PreFine-Homo is highly flexible, capable of delivering homogenization results regardless of the availability of data. Moreover, as the amount of data increases, the accuracy of the operator learned during the pretraining phase of PreFine-Homo improves, leading to a reduction in the number of iterations required during the fine-tuning phase. Therefore, PreFine-Homo represents a novel homogenization computational framework that continuously learns and evolves over time.

\section{Results} \label{sec:Result}
In this section, we evaluate the generalization ability of PreFine-Homo from different aspects, including different geometries, different materials, different resolutions, and extrapolation ability where the test set distribution is entirely different from the training set.

\subsection{Different geometry\label{sec:geo}}

To assess the performance of PreFine-Homo across various geometries, we employed a range of periodic structures known as Triply Periodic Minimal Surfaces (TPMS) detailed in \Cref{sec:TPMS}. TPMS represents a classical type of periodic structure in artificial cellular materials, characterized by its topology-driven properties.

We generated 1800 TPMS unit cells with resolution 128*128*128 (The resolution of the training and testing sets are both 128) for studying the proposed homogenization method. 
These 1800 unit cells are divided into 3 groups ("Sheet-networks": Schoen Gyroid, Schwarz Diamond, and Fischer Kosh S detailed in \Cref{sec:TPMS}).
For each TPMS category, we acquired 600 different geometric structure data sets with varying volume fractions. 
Specifically, the volume fraction of datasets varies from 0.26 to 0.66.
We shuffled all the data to divide it into training and testing sets, ensuring that the test set does not appear in the training set.
These datasets were then processed through a traditional finite element program for homogenization to calculate the corresponding displacement fields and the effective elastic tensor.

The input data were the Poisson's ratios, meaning the 3D matrix contained zeros where there was no material, and the corresponding Poisson's ratio values at material points.  The FNO model's mode is set to 12, width to 32, with a total of 28,322,738 parameters, and a learning rate of 0.001. These parameters can be adjusted according to the hardware capabilities.
Our dataset consists of 150 samples each of 128×128×128 resolution data from three types of "Sheet-networks": Schoen Gyroid, Schwarz Diamond, and Fischer Kosh S. The input has 4 channels, representing Poisson's ratio, and the x, y, and z coordinates. The output consists of 18 displacement fields, the 3 displacement fields under 6 different loading conditions. Currently, we use a GPU with 32GB of memory, so our input data is limited to 450 samples of 128-resolution data. With multi-GPU training, the dataset can be expanded. Despite using a single GPU, our results are already outstanding. The training set consists of 400 samples, while the test set contains 50 samples. Please note that we have combined all types of TPMS in training and test sets to demonstrate the powerful fitting capability of PreFine-Homo.

\begin{figure}
	\begin{centering}
		\includegraphics[scale=0.30]{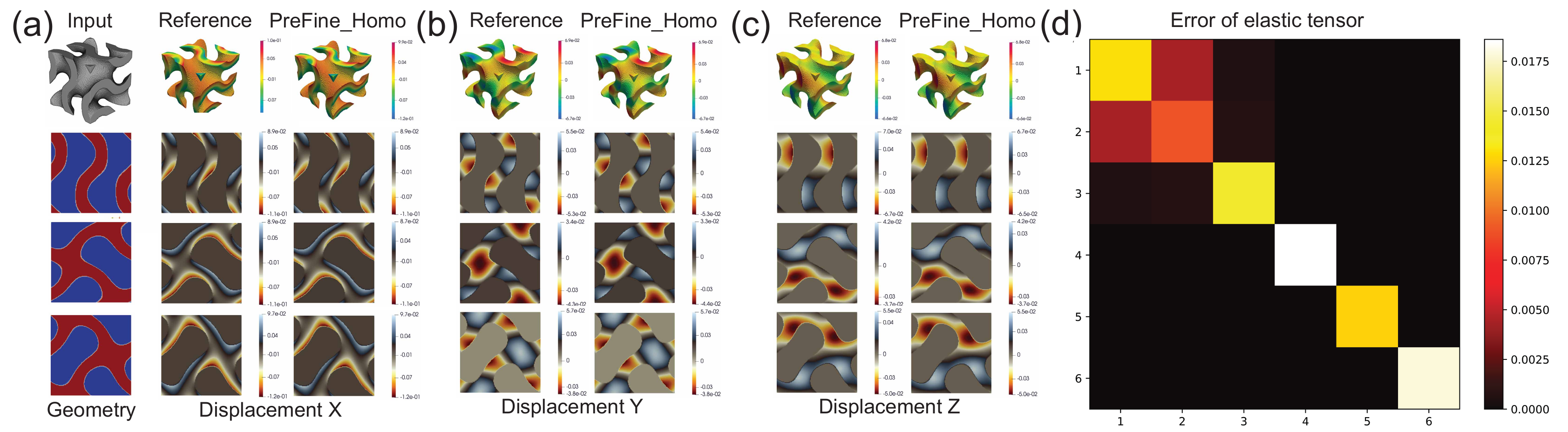}
		\par\end{centering}
	\caption{Prediction results of the proposed PreFine-Homo model in the pretraining phase  across different geometries (Resolution 128) for Schoen Gyroid of "sheet-networks" in the test set under uniaxial tension in the x-direction. (a) The left column shows the geometric input for PreFine-Homo, and the right two columns show the FEM reference solution and the prediction by PreFine-Homo for the $x$-direction displacement field, $y$-direction displacement field(b), $z$-direction displacement field(c).  (d) The relative error of the effective elastic tensor in the corresponding positions by PreFine-Homo, using the traditional finite element method as the reference solution. The second, third, and fourth rows of (a, b, c) show the results for different 2D cross-sections.
		\label{fig:geo_sheet_SG} }
\end{figure}

We first demonstrate the capabilities of PreFine-Homo in the pretraining phase.
\Cref{fig:geo_sheet_SG}a shows geometric data for Schoen Gyroid of "sheet-networks". We kept the Poisson's ratio constant to focus on the PreFine-Homo model's generalization ability across different geometries. The PreFine-Homo in the pretraining phase generated the 18 needed displacement fields for the homogenization process (three displacement fields for each of the six types of loading). 

Due to the extensive amount of model output, we showcased a comparison between PreFine-Homo and the reference solution from FEM under the first type ($x$-direction tension) of loading. \Cref{fig:geo_sheet_SG}a, b, and c respectively show $x$, $y$, and $z$ direction displacement field in the test set. It is evident that PreFine-Homo in the pretraining phase closely matches the FEM reference solution.

After obtaining the displacement fields via PreFine-Homo in the pretraining phase, we integrated these fields into the homogenization process to determine the final effective elastic tensor. It is important to note that PreFine-Homo applies an appropriate scaling factor to the predicted effective elastic tensor based on the training set to further enhance prediction accuracy. The details of the scaling factor are provided in the \Cref{sec:Factor_technique}.

\Cref{fig:geo_sheet_SG}d represents the calculation of relative errors $\mathcal{L}_{ijkl}^{error}$ between the effective elastic tensor computed by PreFine-Homo and FEM (reference solution) in the test set, where the matrix has a 6×6 shape. The relative errors $\mathcal{L}_{ijkl}^{error}$ is calculated as follows:
\begin{equation}
	\mathcal{L}_{ijkl}^{error}=\frac{\vert E_{ijkl}^{P}-E_{ijkl}^{FEM}\vert}{E_{ijkl}^{FEM}},
\end{equation}
where $E_{ijkl}^{P}$ and $E_{ijkl}^{FEM}$ are the predictions from PreFine-Homo and FEM, respectively. The areas in pure black were not included in the relative error calculation because the reference values are close to zero. We can observe that PreFine-Homo predicts the effective elastic tensor with very high accuracy, with a maximum error of less than 2\%.

\begin{figure}
	\begin{centering}
		\includegraphics[scale=0.31]{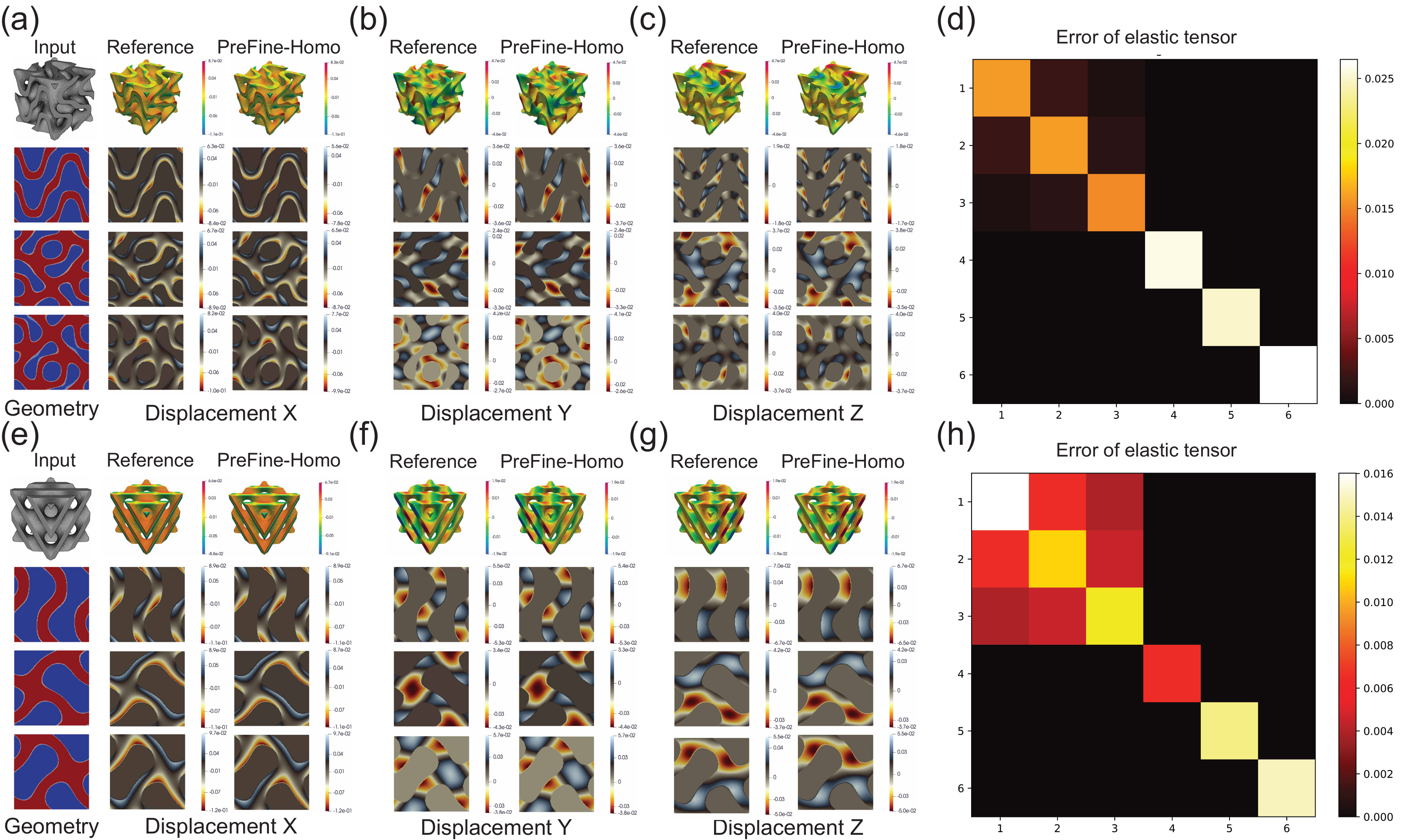}
		\par\end{centering}
	\caption{Prediction results  of the proposed PreFine-Homo model in the pretraining phase across different geometries (Resolution 128) for Schwarz Diamond (a,b,c,d) and Fischer-Koch S (e,f,g,h) of both "sheet-networks" in the test set under uniaxial tension in the x-direction.
		\label{fig:geo_sheet_SD_FKS} }
\end{figure}

To highlight the performance of PreFine-Homo in the pretraining phase on TPMS geometries, \Cref{fig:geo_sheet_SD_FKS} shows the comparisons in the test set with two other common types of TPMS: Schwarz Diamond and Fischer-Koch S of "sheet-networks". PreFine-Homo closely matches the FEM solutions on the test set across all three types of TPMS, with an average relative error of the effective elastic tensor of only about 1.58\%.

\begin{figure}
	\begin{centering}
		\includegraphics[scale=0.38]{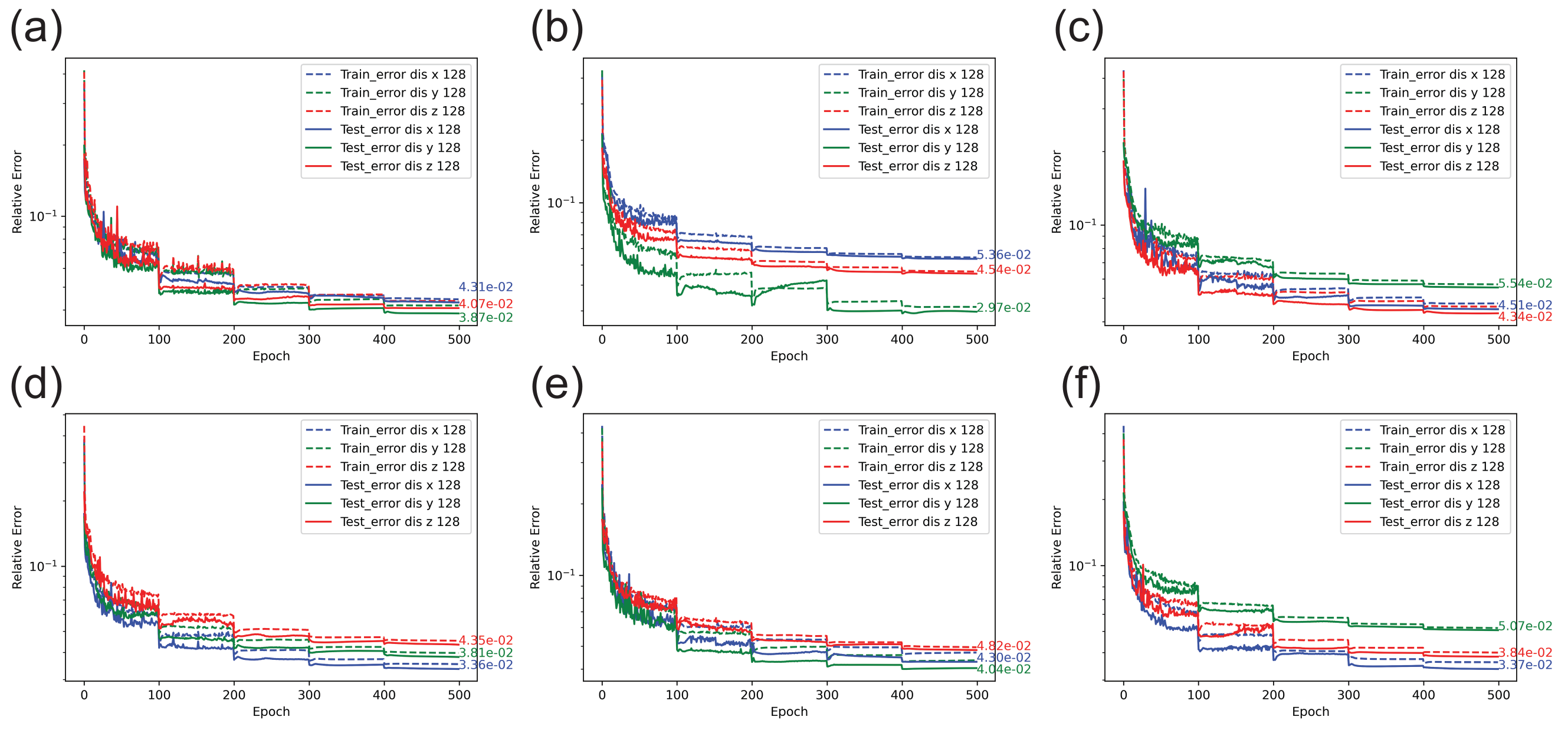}
		\par\end{centering}
	\caption{Training and testing errors of the pretraining phase of the PreFine-Homo for the three displacement fields under six loading conditions in different geometries (Resolution 128). The numbers in the figures represent the final converged values of the testing errors: (a) $x$-direction tension, (b) $y$-direction tension, (c) $z$-direction tension, (d) $xy$-direction shear, (e) $xz$-direction shear, (f) $yz$-direction shear.
		\label{fig:L2_norm_geo} }
\end{figure}

Next, we present the error evolution of FNO in predicting the displacement fields during the homogenization process. \Cref{fig:L2_norm_geo} shows the evolution trends of the training and testing errors. Overall, we found that the testing errors converge to approximately 3\% to 5\%.

Above, we demonstrated the accuracy and error evolution. Now, we focus on the efficiency. The greatest advantage of PreFine-Homo is its efficiency, as it can quickly predict the displacement fields needed for the homogenization process. 

As shown in \Cref{tab:PreFine-Homo_geo_time}, the efficiency gains of PreFine-Homo in the pretraining phase are significantly higher compared to the Finite Element Method (FEM). Since FEM relies on iterative algorithms, its computational time depends heavily on the convergence threshold ($tol$). For instance, when $tol = 10^{-10}$, the average computational time for FEM is approximately 4000 seconds. In contrast, PreFine-Homo in the pretraining phase requires only about 0.43 seconds, demonstrating a remarkable improvement in computational efficiency for displacement field prediction.

Moreover, as illustrated in \Cref{fig:L2_norm_geo}, PreFine-Homo maintains high accuracy, with testing errors converging to approximately 3\% to 5\%. This combination of efficiency and precision makes PreFine-Homo a powerful tool for large-scale homogenization tasks, significantly outperforming traditional methods in both speed and reliability.

The key to improving homogenization efficiency is to accelerate the prediction of the displacement field required in homogenization. The core reason that PreFine-Homo is faster than traditional algorithms lies in the utilization of operator learning to boost the efficiency of displacement field prediction, thereby enhancing the overall homogenization efficiency.

Next, we demonstrate the capabilities of PreFine-Homo in the fine-tuning phase. The core idea of PreFine-Homo in the fine-tuning phase is to use the predictions from the pretraining phase as the initial solution for the iterative algorithm. It is worth noting that the fine-tuning phase of PreFine-Homo is optional. If higher precision is not required and efficiency is the primary concern, one can skip the fine-tuning phase and rely solely on the pretraining phase of PreFine-Homo.

A key advantage of PreFine-Homo in the fine-tuning phase lies in its ability to significantly reduce the number of iterations required for convergence compared to traditional PDE solvers like the Finite Element Method (FEM). This reduction is achieved by leveraging the high-quality initial solution provided by the pretraining phase of PreFine-Homo. By starting with an initial solution that is already close to the true solution, PreFine-Homo in the fine-tuning phase can achieve convergence with far fewer iterations, thereby enhancing computational efficiency while maintaining high accuracy as same as traditional PDE solvers.

As shown in \Cref{tab:PreFine-Homo_geo_time}, the iteration counts for PreFine-Homo in the fine-tuning phase are substantially lower than those of FEM across various convergence thresholds ($tol$). The iteration counts are statistically averaged across the entire test set, which includes 50 data points from three types of TPMS structures. For example, at $tol = 10^{-5}$, FEM requires an average of 238.1, 203.6, and 193.8 iterations for Schoen Gyroid (SG), Schwarz Diamond (SD), and Fischer-Koch S (FKS) geometries, respectively. In contrast, PreFine-Homo\_Fine requires only 84.28, 70.11, and 65.16 iterations for the above geometries, representing a reduction of approximately 65\%, 66\%, and 66\%, respectively.

\begin{figure}
	\begin{centering}
		\includegraphics[scale=0.6]{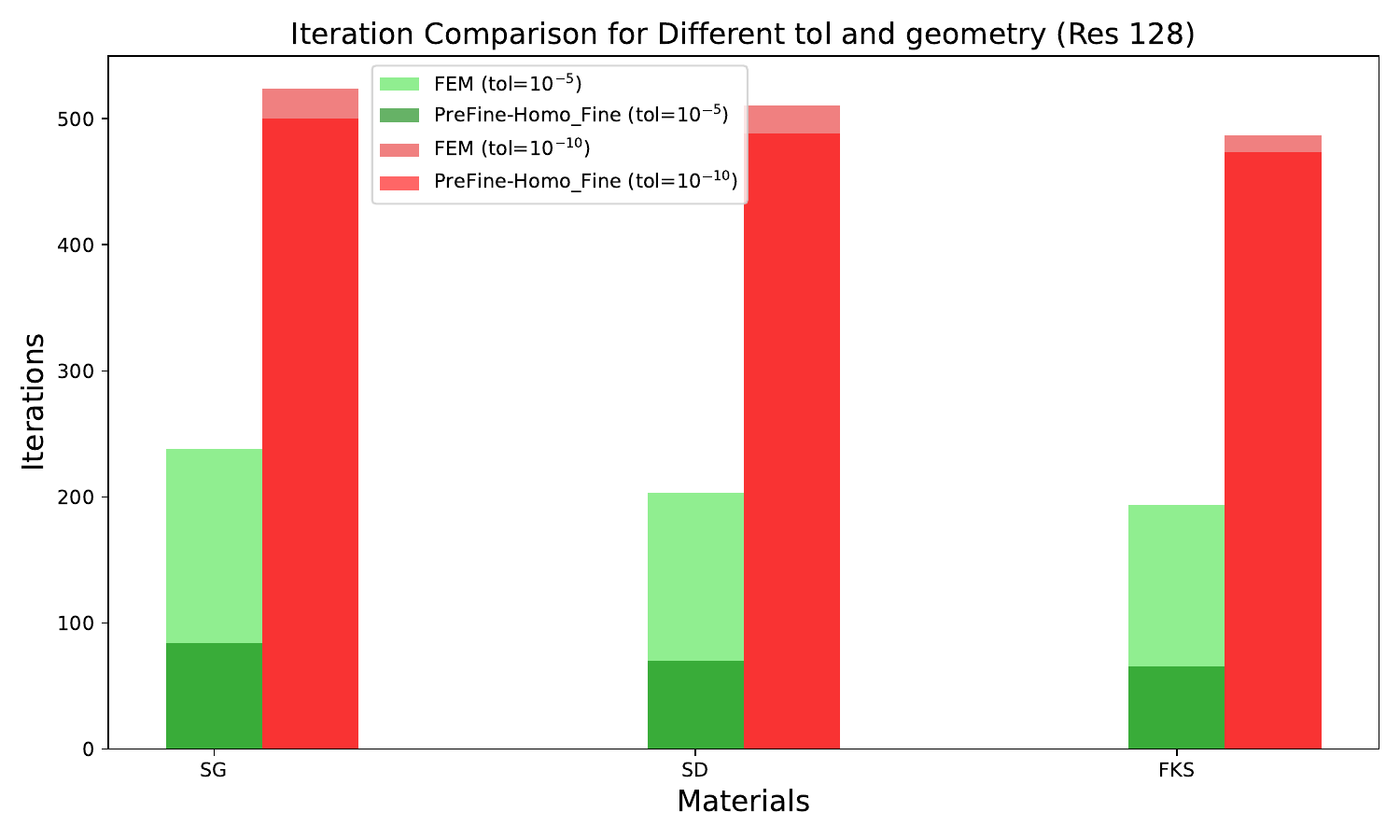}
		\par\end{centering}
	\caption{Comparison of the number of iterations required between the traditional PDE solver FEM and the fine-tuning phase of PreFine-Homo across different tolerances (tol) and all different geometry test cases (a single test case) in the iterative algorithm.
		\label{fig:geo_iterations} }
\end{figure}

This trend holds even for stricter convergence thresholds. At $tol = 10^{-10}$, FEM requires 523.6, 510.7, and 487.2 iterations for SG, SD, and FKS, respectively, while PreFine-Homo\_Fine requires 500.2, 488.2, and 473.4 iterations. Although the reduction in iteration counts is less pronounced at higher precision levels as shown in \Cref{fig:geo_iterations}, this is because the benefits of a good initial solution diminish as $tol$ decreases, as detailed in \Cref{sec:benifits_fine_tuning}. The increased number of iterations required for convergence at higher precision levels tends to overshadow the early-stage benefits provided by the initial solution. Nevertheless, PreFine-Homo still demonstrates a clear advantage, particularly when combined with its significantly faster pretraining phase.

The reduced iteration counts in the fine-tuning phase of PreFine-Homo directly translate to improved computational efficiency, making it a powerful tool for large-scale 3D elasticity problems. It is important to emphasize that the fine-tuning phase of PreFine-Homo ensures accuracy comparable to that of traditional high-precision PDE solvers. As a result, the error in the effective elastic tensor can approach zero. Therefore, we do not present the homogenization results for the effective elastic tensor here, as its accuracy is entirely determined by the precision of the displacement field.

\begin{table}
	\caption{The efficiency of PreFine-Homo on an average over all different geometry test cases (a single test case) with a resolution of 128. $tol$ is the convergence threshold for the iterative algorithm. FEM represents the average time and average number of iterations per load case for a traditional PDEs solver. PreFine-Homo\_Pre denotes the average time for testing a single case in the pretraining phase of PreFine-Homo. PreFine-Homo\_Fine represents the average time and average number of iterations per load case for testing a single case in the fine-tuning phase of PreFine-Homo. SG, SD, and FKS refer to Schoen Gyroid, Schwarz Diamond, and Fischer-Koch S of "sheet-networks," respectively. \label{tab:PreFine-Homo_geo_time}}
	\begin{adjustbox}{max width=\textwidth}
		\centering
		\begin{tabular}{cccccccc}
			\toprule 
			\multirow{2}{*}{} & \multicolumn{3}{c}{$tol=10^{-5}$} &  & \multicolumn{3}{c}{$tol=10^{-10}$}\tabularnewline
			\cmidrule{2-4} \cmidrule{3-4} \cmidrule{4-4} \cmidrule{6-8} \cmidrule{7-8} \cmidrule{8-8} 
			& SG & SD & FKS &  & SG & SD & FKS\tabularnewline
			\midrule
			FEM:sec/iters & 1809/238.1 & 1726/203.6 & 1592/193.8 &  & 3687/523.6 & 4757/510.7 & 3558/487.2\tabularnewline
			\midrule 
			PreFine-Homo\_Pre:sec & 0.42365 & 0.43983 & 0.44476 &  & 0.42365 & 0.43983 & 0.44476\tabularnewline
			\midrule 
			PreFine-Homo\_Fine:sec/iters & \textbf{867.4/84.28} & \textbf{736.3/70.11} & \textbf{681.1/65.16} &  & \textbf{3504/500.2} & \textbf{4514/488.2} & \textbf{3276/473.4}\tabularnewline
			\bottomrule
		\end{tabular}
	\end{adjustbox}
\end{table}

To further explore the performance of PreFine-Homo, our next numerical experiments will test PreFine-Homo's performance with inputs of different materials and geometries simultaneously, evaluating how well PreFine-Homo performs.

\subsection{Different material\label{sec:diff_material}}

Unlike the geometric experiments mentioned before, this experiment modifies the input Poisson's ratio to account for different materials. The decision not to consider varying Young's modulus because that final effective elastic tensor can be linearly adjusted by simply multiplying by the Young's modulus at the end. Therefore, in different materials, we only need to consider different Poisson ratios. 

It should be noted that the Poisson's ratio is uniform across all spatial points, but it varies between different datasets. If the Poisson's ratio were to vary spatially, the amount of data required would increase significantly. In this preliminary work, we do not consider spatial variations in the Poisson's ratio. In the future, we believe that our proposed PreFine-Homo in the pretraining phase has the potential to handle homogenization problems with spatially varying Poisson's ratios, provided that sufficient data is available.
However, the fine-tuning phase of PreFine-Homo can also ensure successful extrapolation for spatially varying Poisson's ratios, detailed in \Cref{sec:Extra}.

For this experimental dataset, we expanded the variety of TPMS to six types (Schoen Gyroid: "Solid-networks" and "Sheet-networks", Schwarz Diamond: "Solid-networks" and "Sheet-networks", and Fischer Kosh S: "Solid-networks" and "Sheet-networks"). Poisson's ratios range from 0.1 to 0.4, totaling 3600 datasets (600 per type) of 64x64x64 resolution. Here, the geometric volume fractions range from 0.26 to 0.66, and Poisson's ratios range from 0.1 to 0.4. Both volume fractions and Poisson's ratios are uniformly distributed. The dataset comprises 3600 samples in total, with 3000 used for training and 600 for testing, ensuring that the training and testing sets are distinct.

Due to the increased material matrix, our dataset has further expanded. Consequently, for this experiment, we reduced the resolution from 128 (used in previous geometric experiments \Cref{sec:geo}) to 64. Please note that we have combined all types of TPMS in training and test sets to demonstrate the powerful generalizability of PreFine-Homo.

The hyperparameters for the FNO model are as follows: the number of modes is set to 30, the width is set to 32, with a total of 442,379,186 parameters,  and the learning rate is 0.001. Model parameters are adjusted according to the available hardware capabilities.

\begin{figure}
	\begin{centering}
		\includegraphics[scale=0.38]{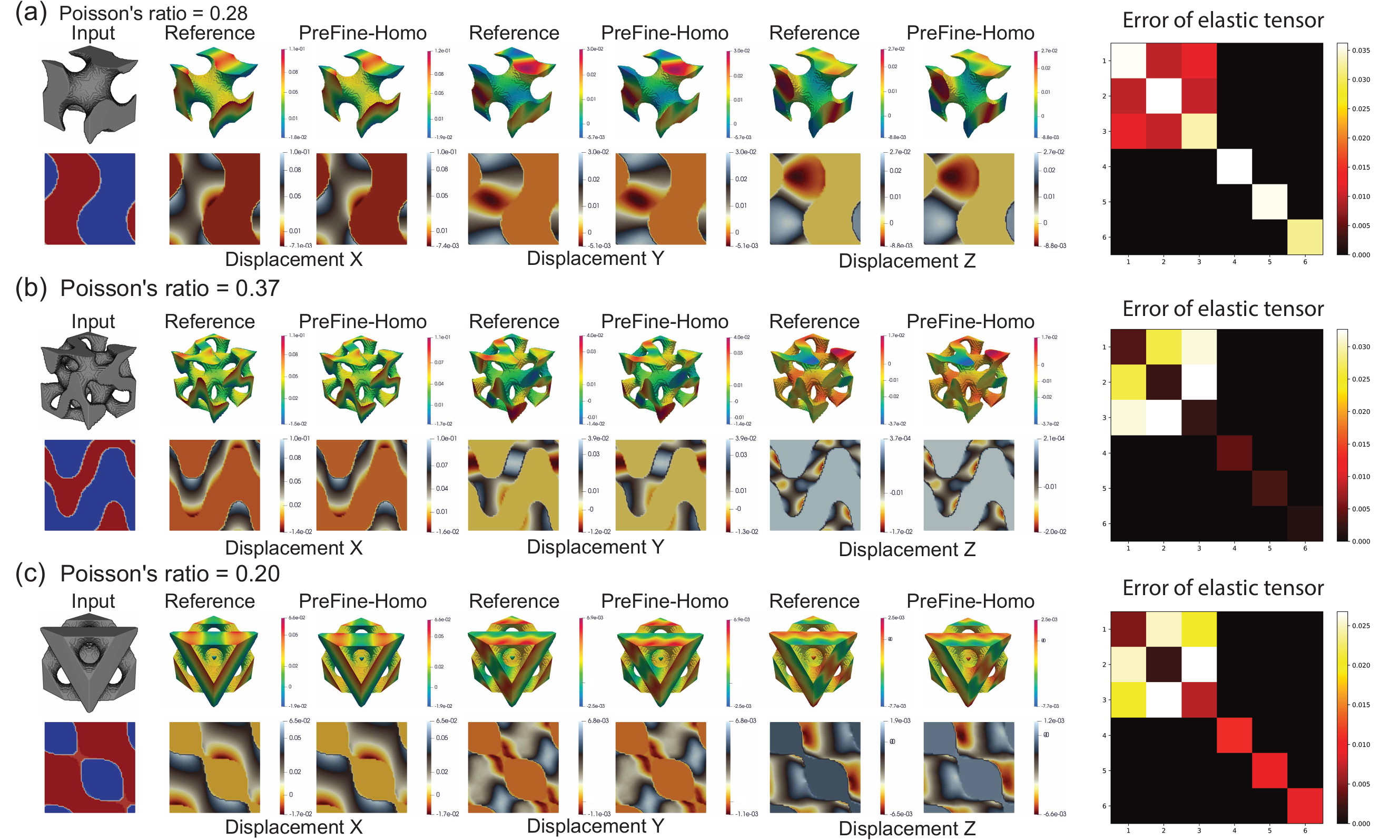}
		\par
	\end{centering}
	\caption{Prediction results of the proposed PreFine-Homo model in the pretraining phase  across different geometries (Resolution 64) and materials (Poisson's ratio and Young's modulus) in "Solid-networks" on the test set under uniaxial tension in the x-direction. (a) Schoen Gyroid (b) Schwarz Diamond (c) Fischer Kosh S.
		\label{fig:material_solid} }
\end{figure}

\begin{figure}
	\begin{centering}
		\includegraphics[scale=0.38]{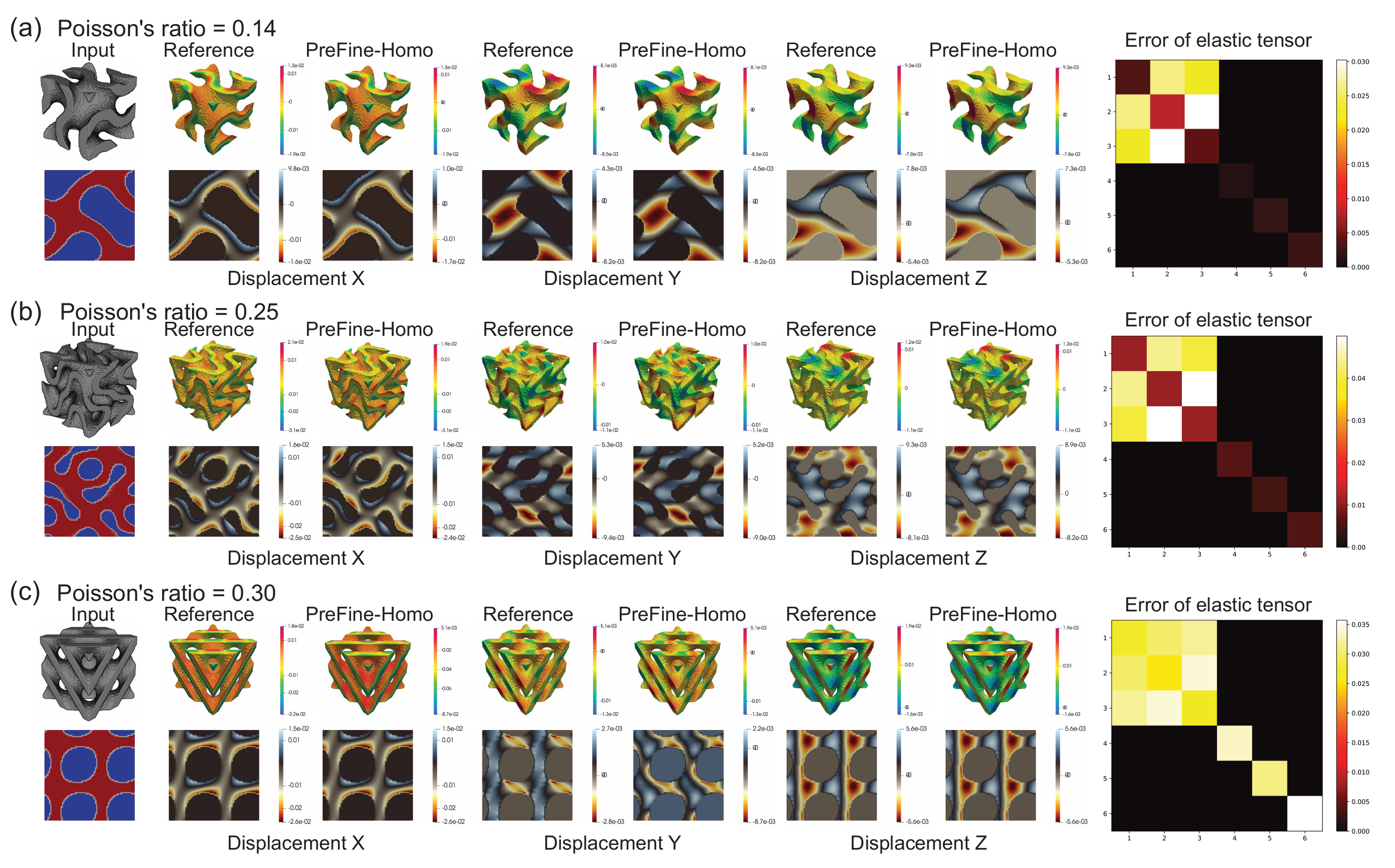}
		\par
	\end{centering}
	\caption{Prediction results of the proposed PreFine-Homo model in the pretraining phase  across different geometries (Resolution 64) and materials (Poisson's ratio and Young's modulus) in "Sheet-networks" on the test set under uniaxial tension in the x-direction. (a) Schoen Gyroid (b) Schwarz Diamond (c) Fischer Kosh S.
		\label{fig:material_sheet} }
\end{figure}

\Cref{fig:material_solid} and \Cref{fig:material_sheet} respectively show the results for "Solid-networks" and "Sheet-networks" in the test set under uniaxial tension in the x-direction. Our results reveal that the PreFine-Homo in the pretraining phase achieved impressive generalization performance across different Poisson ratios and geometries. 

When incorporated into the conventional homogenization program, the maximal relative error in the final effective elastic tensor is less than 5\% and the average relative error is 1.60\%, indicating the high accuracy and super efficiency of our proposed PreFine-Homo in the pretraining phase.

\begin{figure}
	\begin{centering}
		\includegraphics[scale=0.36]{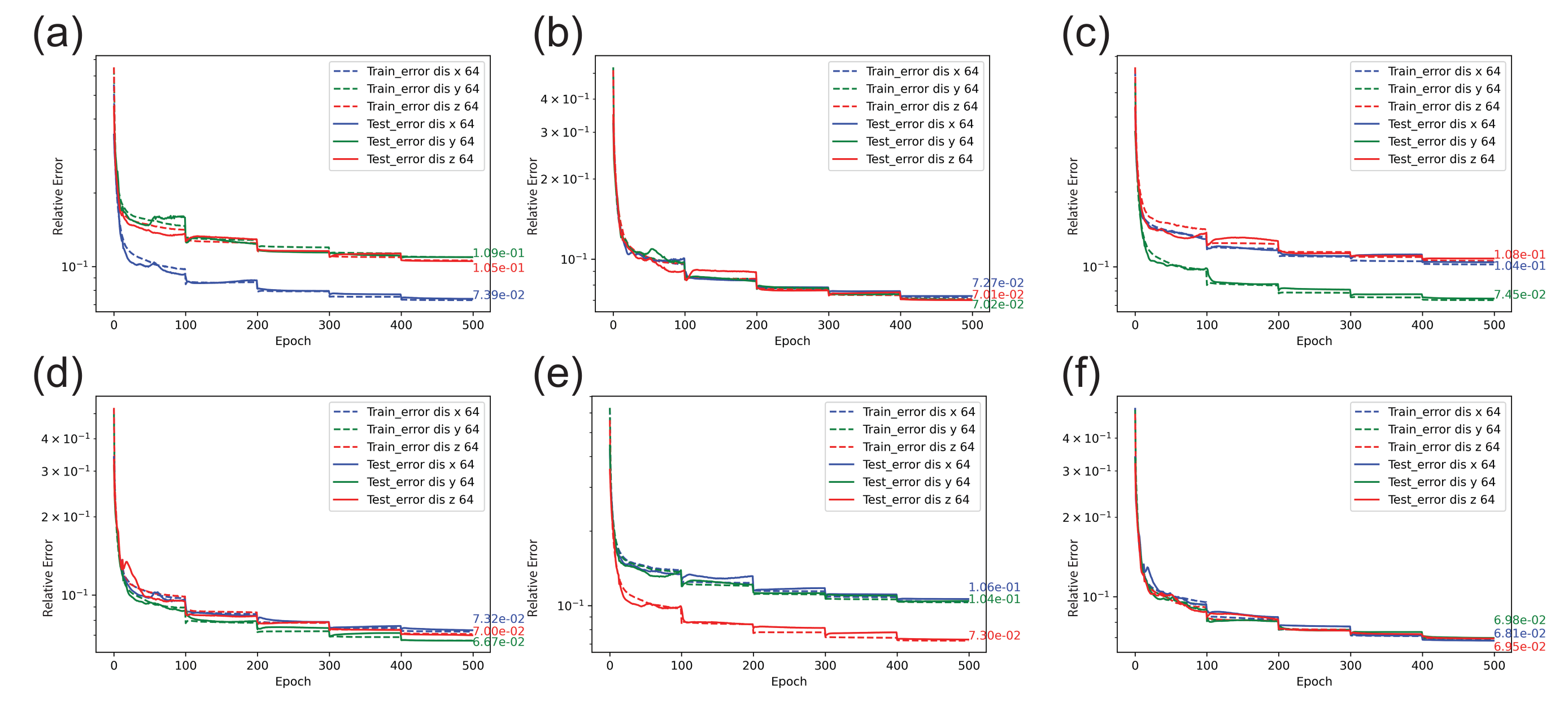}
		\par\end{centering}
	\caption{Training and testing errors of the pretraining phase of of PreFine-Homo for the three displacement fields under six loading conditions in different geometries (Resolution 64) and materials (Poisson's ratio and Young's modulus). The numbers in the figures represent the final converged values of the testing errors: (a) $x$-direction tension, (b) $y$-direction tension, (c) $z$-direction tension, (d) $xy$-direction shear, (e) $xz$-direction shear, (f) $yz$-direction shear.
		\label{fig:L2_norm_material} }
\end{figure}

\Cref{fig:L2_norm_material} shows the evolution of PreFine-Homo's training and validation errors over epochs  in the pretraining phase. The testing error for the displacement field ranges between 6\% and 10\%. This indicates that PreFine-Homo  in the pretraining phase can predict highly accurate displacement fields across different geometries and materials.

\begin{figure}
	\begin{centering}
		\includegraphics[scale=0.6]{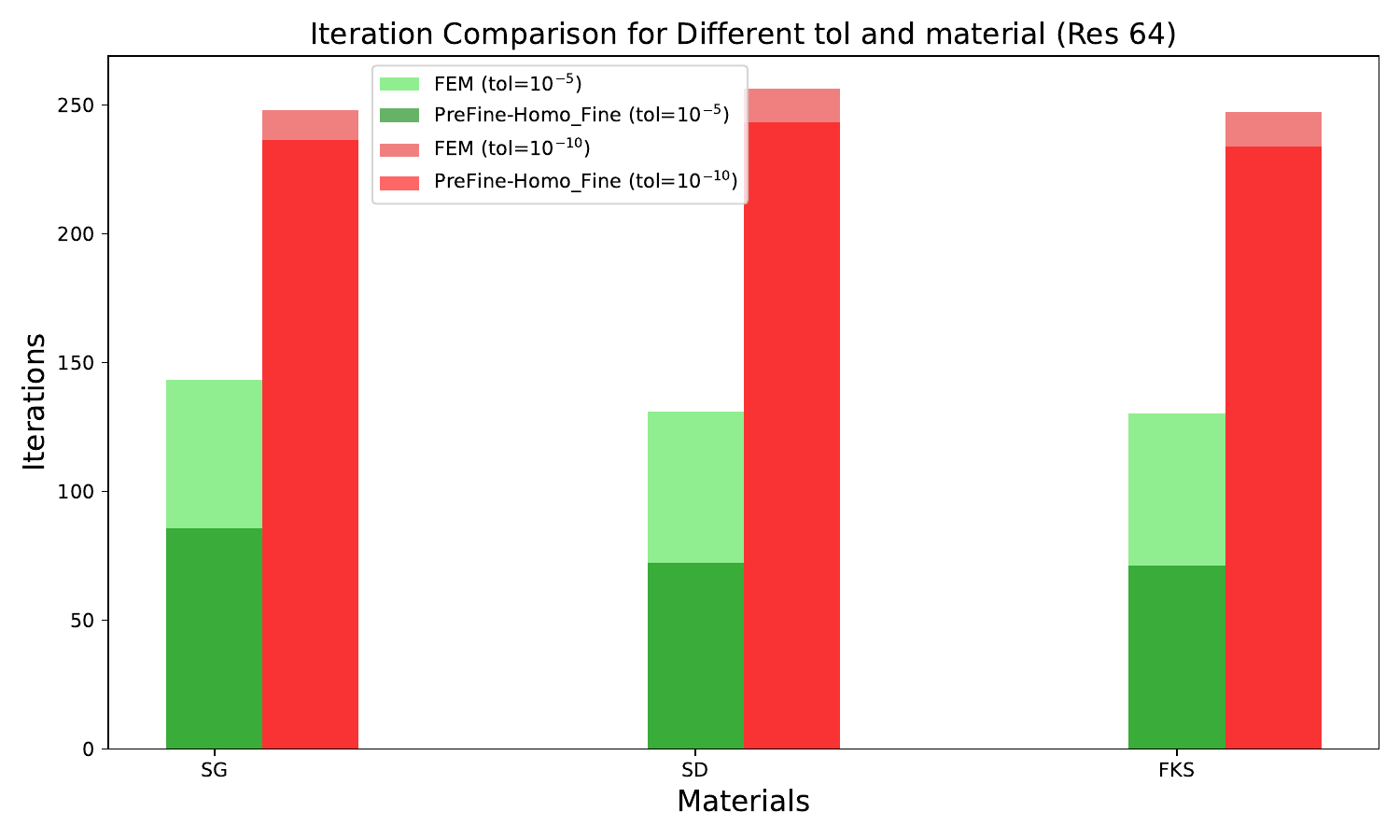}
		\par\end{centering}
	\caption{Comparison of the number of iterations required between the traditional PDEs solver FEM and the fine-tuning phase of PreFine-Homo across different tolerances (tol) and all different material test cases (a single test case) in the iterative algorithm.
		\label{fig:mater_iterations} }
\end{figure}

Next, we demonstrate the capabilities of PreFine-Homo in the fine-tuning phase. As shown in \Cref{tab:PreFine-Homo_material_time} and \Cref{fig:mater_iterations}, the iteration counts for PreFine-Homo are significantly lower than those of FEM across various convergence thresholds ($tol$). The results are averaged over 600 test cases from three TPMS structures. For example, at $tol = 10^{-5}$, PreFine-Homo reduces iteration counts by approximately 40\%–45\% compared to FEM. Even at stricter thresholds like $tol = 10^{-10}$, PreFine-Homo maintains a clear advantage, though the reduction in iterations is less pronounced due to the diminishing benefits of a good initial solution at higher precision levels.

The fine-tuning phase of PreFine-Homo ensures accuracy comparable to traditional high-precision solvers, with errors in the effective elastic tensor approaching zero. Since the accuracy of the elastic tensor is entirely determined by the displacement field, we omit its results here. This efficiency, combined with the pretraining phase's speed, makes PreFine-Homo a powerful tool for large-scale 3D elasticity problems. 

It is worth noting that the FNO model (used in the pretraining phase of PreFine-Homo) in the section on different materials (\Cref{sec:diff_material}) has significantly more parameters (442,379,186) compared to the FNO model in the section on different geometries (\Cref{sec:geo}) (28,322,738). However, the inference time of PreFine-Homo in the pretraining phase is shorter for different materials (Geometry: 0.44 seconds; Material: 0.08 seconds). This is because the inference time of the model depends not only on the number of parameters in the FNO but also on the size of the input and output resolutions. In the section on different geometries, the resolutions of the input and output are both 128, while in the section on different materials, the resolutions are 64.

\begin{table}
	\caption{The efficiency of PreFine-Homo on an average over all different material test cases (a single test case) with a resolution of 64. $tol$ is the convergence threshold for the iterative algorithm. FEM represents the average time and average number of iterations per load case for a traditional PDEs solver. PreFine-Homo\_Pre denotes the average time for testing a single case in the pretraining phase of PreFine-Homo. PreFine-Homo\_Fine represents the average time and average number of iterations per load case for testing a single case in the fine-tuning phase of PreFine-Homo. SG, SD, and FKS refer to Schoen Gyroid, Schwarz Diamond, and Fischer-Koch S of "solid-networks" and "sheet-networks," respectively. \label{tab:PreFine-Homo_material_time}}
	
	\centering
	\begin{adjustbox}{max width=\textwidth}
		\begin{tabular}{cccccccc}
			\toprule 
			\multirow{2}{*} & \multicolumn{3}{c}{$tol=10^{-5}$} &  & \multicolumn{3}{c}{$tol=10^{-10}$}\tabularnewline
			\cmidrule{2-4} \cmidrule{3-4} \cmidrule{4-4} \cmidrule{6-8} \cmidrule{7-8} \cmidrule{8-8} 
			& SG & SD & FKS &  & SG & SD & FKS\tabularnewline
			\midrule
			FEM:sec/iters & 153.7/143.3 & 152.7/130.8 & 145.6/130.2 &  & 249.7/248.1 & 275.1/256.2 & 245.1/247.1\tabularnewline
			\midrule 
			PreFine-Homo\_Pre:sec & 0.088000 & 0.088999 & 0.087580 &  & 0.088000 & 0.088999 & 0.087580\tabularnewline
			\midrule 
			PreFine-Homo\_Fine:sec/iters & \textbf{103.5/85.76} & \textbf{96.96/72.18} & \textbf{92.15/71.37} &  & \textbf{234.2/236.3} & \textbf{264.5/243.4} & \textbf{243.5/234.0}\tabularnewline
			\bottomrule
		\end{tabular}
	\end{adjustbox}
\end{table}

\subsection{Different resolution}
The experiments involving different resolutions refer to training the model at lower resolutions and then testing it at higher resolutions. We conducted training with datasets at resolutions of 32×32×32, 64×64×64, and 128×128×128, and performed tests on all at a resolution of 128×128×128 to evaluate the model's cross-resolution performance in the pretraining phase.

\begin{figure}
	\begin{centering}
		\includegraphics[scale=0.45]{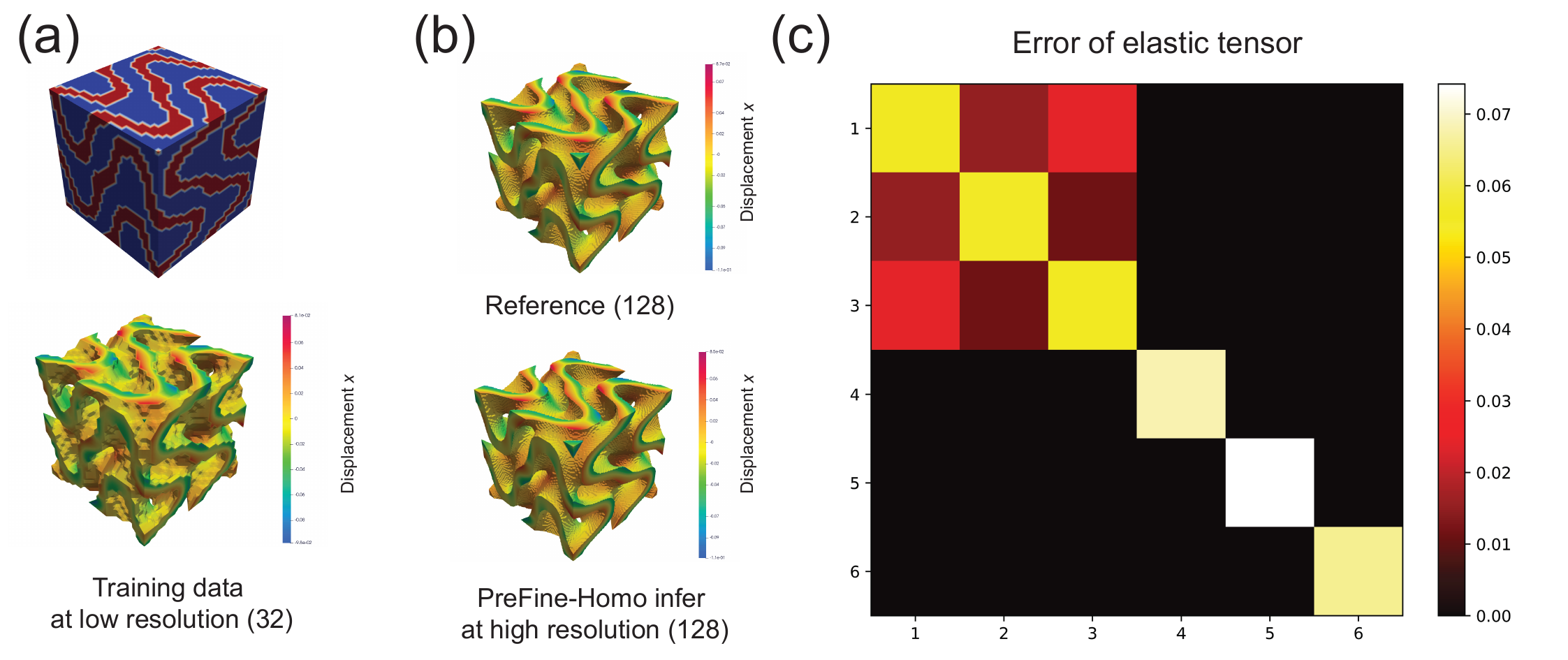}
		\par\end{centering}
	\caption{ Results of PreFine-Homo in the pretraining phase  across different resolutions: (a) Data with a resolution of 32 used as the training set for PreFine-Homo, (b) Reference solution and the prediction (displacement $x$) by PreFine-Homo with a resolution of 128 in the test set under uniaxial tension in the x-direction, (c) The relative error of the effective elastic modulus tensor in the test set.
		\label{fig:diff_res} }
\end{figure}

\Cref{fig:diff_res}a shows the geometry and corresponding displacement fields for the training set with a resolution of 32. Note that although our training set is at a resolution of 32, we perform predictions at a resolution of 128. \Cref{fig:diff_res}b demonstrates PreFine-Homo's outstanding cross-resolution capability in the pretraining phase. \Cref{fig:diff_res}c shows the relative error of the final predicted effective elastic tensor. The results indicate that despite the low quality of the training set, the pretraining phase of PreFine-Homo performs excellently on the high-resolution test set. \Cref{tab:PreFine-Homo_performance} shows that the relative test errors for the effective elastic tensor are 15.37\% and 1.58\% for training sets at resolutions of 32 and 128, respectively.

\begin{table}
	\caption{The comparison between PreFine-Homo and the traditional numerical homogenization method (Finite Element Method, FEM) in terms of accuracy and efficiency is conducted based on different training resolutions (32, 64, 128) and the same testing resolution (128). The time reported for FEM (in seconds) refers to the average computational time of FEM at a resolution of 128 when the convergence threshold of the iterative algorithm is set to $tol = 10^{-10}$. The time reported for PreFine-Homo in the table specifically refers to the pretraining phase of PreFine-Homo, which handles all testing cases at a resolution of 128, even though the training is performed at different resolutions (32, 64, 128).}
	
	\begin{center}
		\begin{adjustbox}{max width=\textwidth}
			\begin{tabular}{cccccc}
				\toprule 
				Train Res. & Time: FEM (s)  & Time: PreFine-Homo (s) & Dis: Train error & Dis: Test error & Effective modulus error\tabularnewline
				\midrule 
				32*32*32 & \multirow{3}{*}{4010} & \multirow{3}{*}{\textbf{0.43776}} & 0.0578 & 0.2133 & 0.1537\tabularnewline
				64*64*64 &  &  & 0.0860 & \multirow{1}{*}{0.1347} & 0.1074\tabularnewline
				128*128*128 &  &  & 0.0747 & \textbf{0.0461} & \textbf{0.0158}\tabularnewline
				\bottomrule
			\end{tabular}
		\end{adjustbox}
		\par\end{center}
	
	\label{tab:PreFine-Homo_performance}
\end{table}

\begin{figure}
	\begin{centering}
		\includegraphics[scale=0.40]{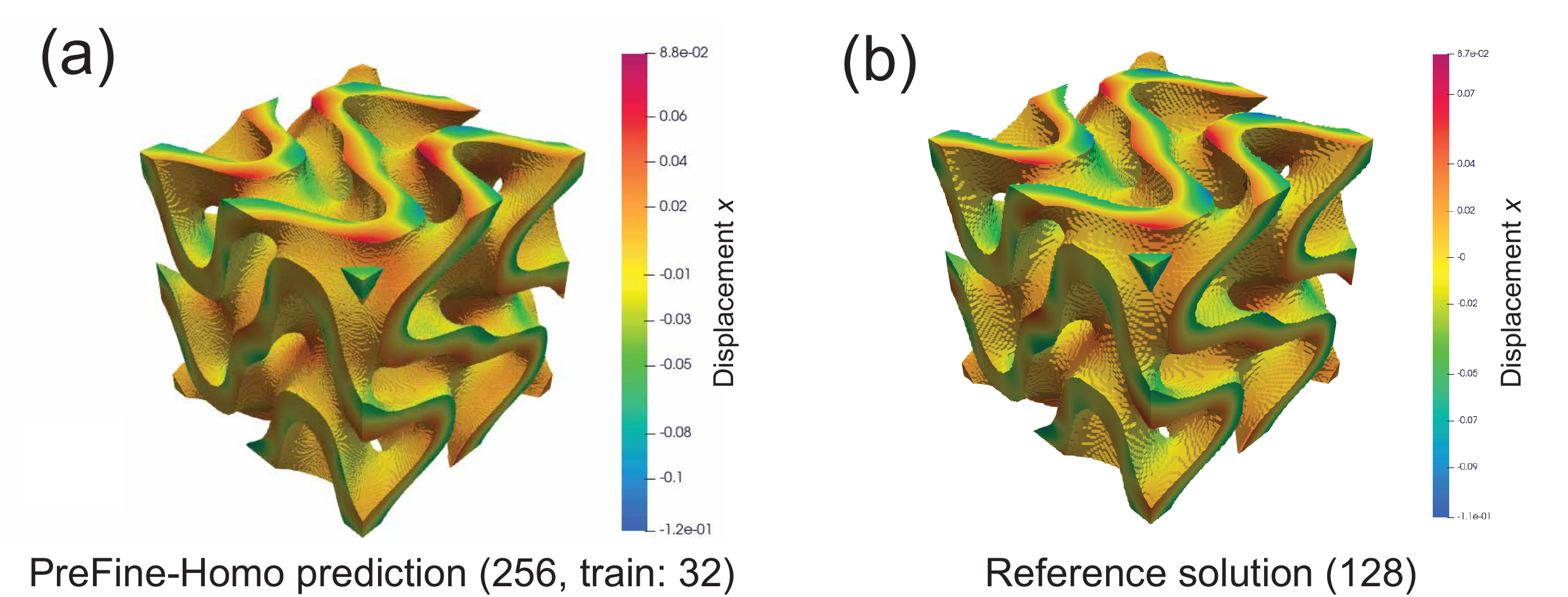}
		\par\end{centering}
	\caption{Super resolution results of the pretraining phase of PreFine-Homo in the test set under uniaxial tension in the x-direction: (a) Prediction (displacement $x$) by PreFine-Homo at a resolution of 256, (b) Reference solution (displacement $x$) at a resolution of 128, the traditional FEM cannot handle the resolution of 256 (192G memory exceeded).
		\label{fig:super_res} }
\end{figure}

Owing to PreFine-Homo's capacity to operate across resolutions, it theoretically can predict outcomes at any resolution. Thus, we tested PreFine-Homo's capabilities at an ultra-high resolution (256), with results depicted in \Cref{fig:super_res}. At a resolution of 256, compared to predictions at a resolution of 128, PreFine-Homo demonstrates remarkable robustness. Notably, FEM cannot handle the resolution of 256 on a system with 192GB of memory due to memory limitations. Even with a larger memory CPU, FEM would take over 10 hours to compute at a resolution of 256, whereas PreFine-Homo requires merely 20 seconds. Since the resolution of 128 has already converged, we use the resolution of 128 as the reference solution in this context.

Next, we demonstrate the performance of PreFine-Homo in the fine-tuning phase under different training resolutions. During the pretraining phase of PreFine-Homo, the training data resolutions are set to 32, 64, and 128, respectively. However, the resolution for testing during the fine-tuning phase is consistently 128. \Cref{tab:PreFine-Homo_performance} shows the errors of PreFine-Homo in the pretraining phase for different training resolutions. It is evident that lower training resolutions result in larger test errors at the testing resolution of 128.

\begin{figure}
	\begin{centering}
		\includegraphics[scale=0.60]{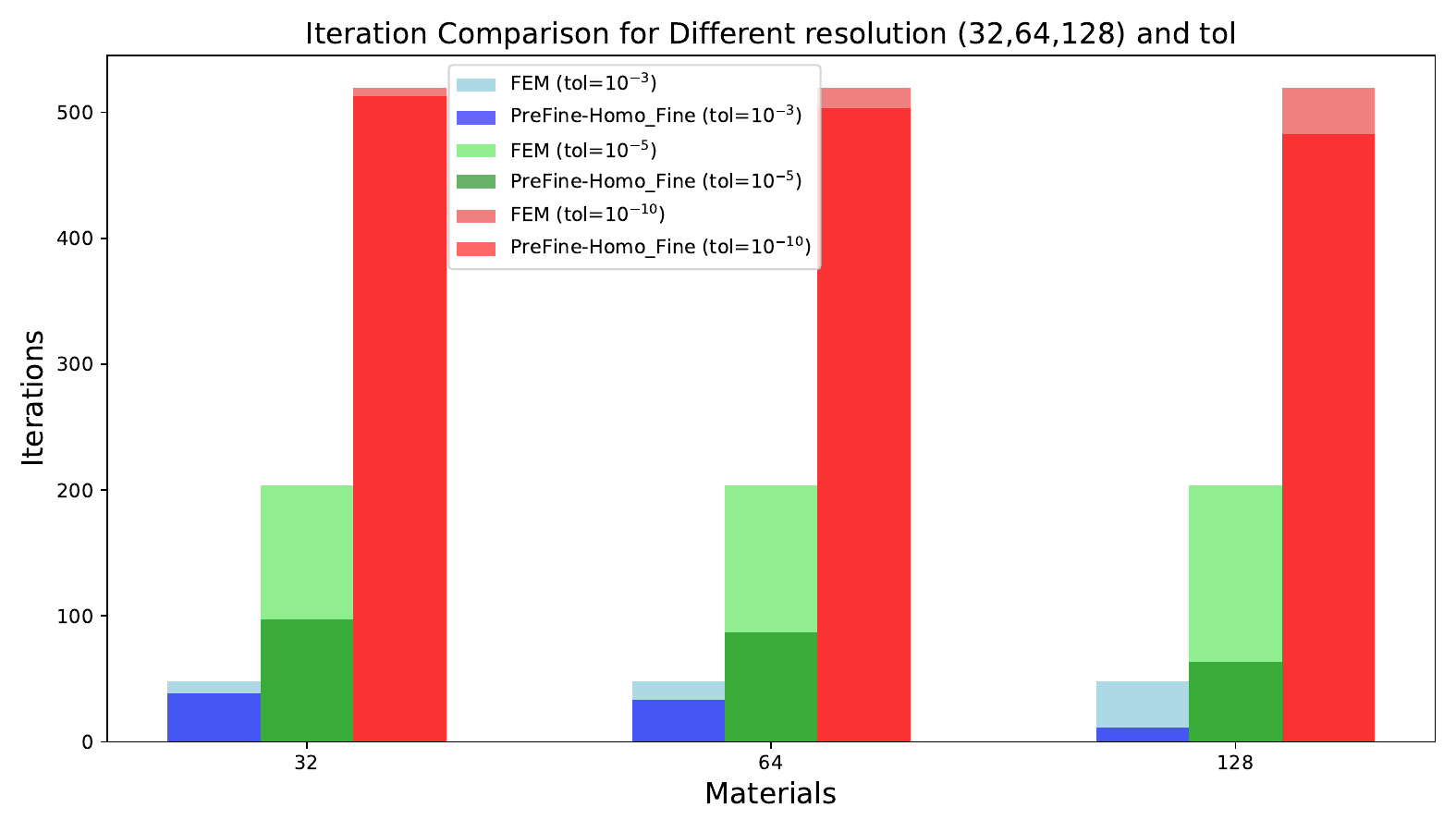}
		\par\end{centering}
	\caption{Comparison of the number of iterations required between the traditional PDEs solver FEM and the fine-tuning phase of PreFine-Homo across different tolerances (tol) and resolution (Training resolutions: 32, 64, 128; Testing resolutions: 128).
		\label{fig:res_iterations} }
\end{figure}

Naturally, we are curious about the performance of different pretrained models of PreFine-Homo and how the number of iterations in the fine-tuning phase changes. \Cref{fig:res_iterations} compares the number of iterations required in the fine-tuning phase of PreFine-Homo with those of the traditional PDEs solver (FEM) for different training resolutions. It is clear that, across various convergence thresholds ($tol$), the more accurate the initial solution provided by the pretraining phase of PreFine-Homo, the fewer iterations are required in the fine-tuning phase.

Therefore, in the future, as the amount of data increases, the pretraining phase of PreFine-Homo will become increasingly accurate, significantly reducing the number of iterations required in the fine-tuning phase. PreFine-Homo is a self-learning model with the potential for continuous improvement, as detailed in \Cref{sec:self_learning}.

To sum up, the model exhibits commendable performance across different resolutions. This demonstrates that PreFine-Homo can integrate data of various resolutions, which is crucial in practical applications as we can use all available resolution data to train PreFine-Homo. Furthermore, if data at lower resolutions contain errors, it could diminish the model's performance. Therefore, exploring how to better train models, especially when lower-resolution data contain errors, is crucial for future research.

\subsection{Extrapolation ability\label{sec:Extra}}

In this section, we demonstrate the extrapolation ability of PreFine-Homo.
We experimented with training on three types of TPMS within only "Sheet-networks" (Schoen Gyroid, Schwarz Diamond, and Fischer Kosh S) but tested on the same three types of TPMS in "Solid-networks" during the pretraining phase of PreFine-Homo. The pretraining phase of PreFine-Homo achieved impressive results in this scenario. However, when testing on structures generated by Gaussian Random Fields (GRF), the pretraining phase of PreFine-Homo showed limitations. Note that the pretraining phase of PreFine-Homo uses a model trained on TPMS geometries with a resolution of 128, as detailed in \Cref{sec:geo}.  The porous structures are generated by constructing a Gaussian random field through the superposition of multiple sinusoidal waves, followed by applying a threshold to obtain three-dimensional porous structures with varying porosities. For more details on the generation by GRF, please refer to the method proposed by \citep{soyarslan20183d}.

\begin{figure}
	\centering
	\includegraphics[scale=0.50]{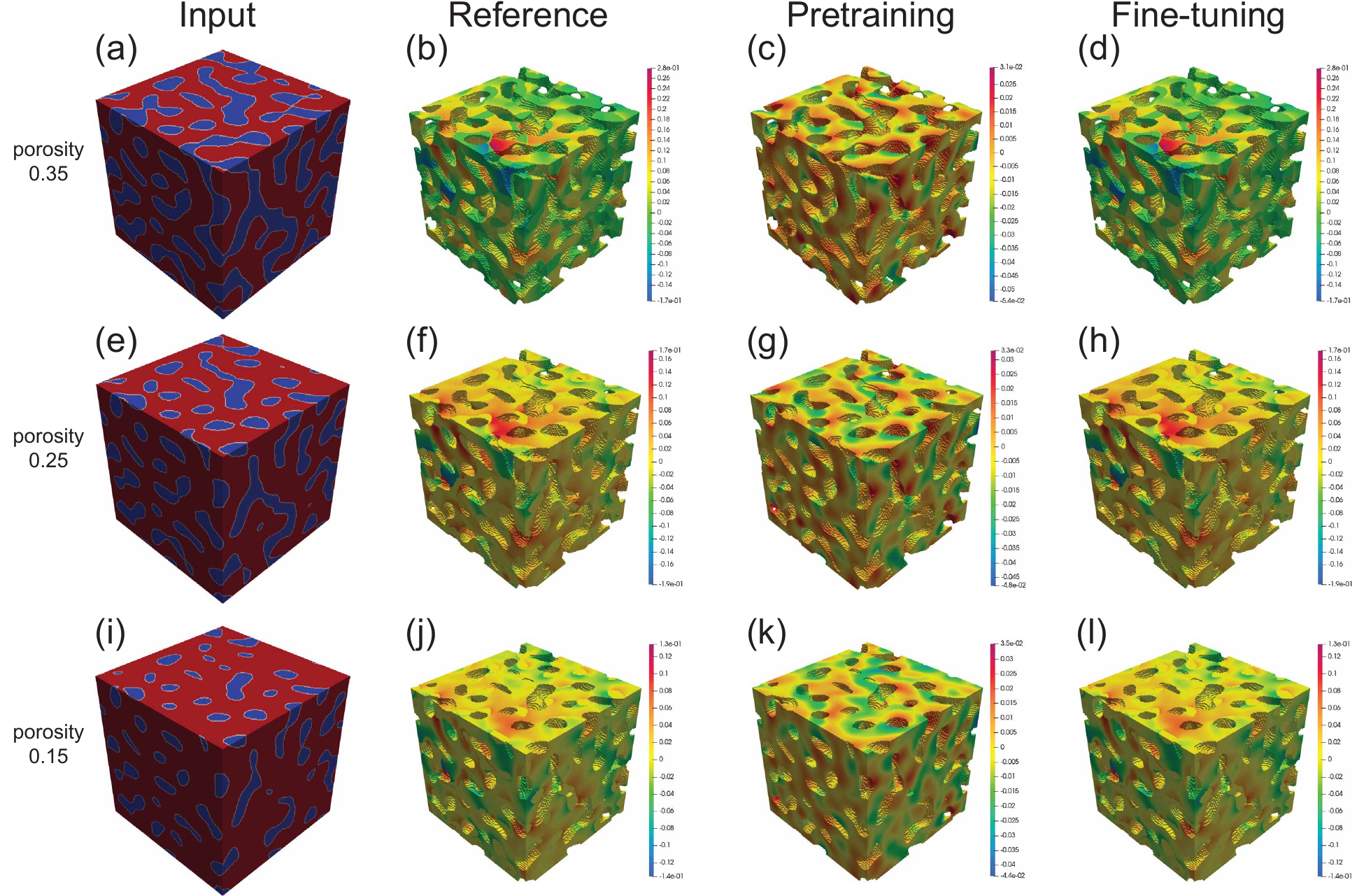}
	\caption{Prediction results of PreFine-Homo on data generated by GRF (Gaussian Random Field) in the test set under uniaxial tension in the x-direction: The first column (a,e,i) shows the geometry and topology of GRF, where blue represents holes and red represents material; the second column (b,f,j) shows the FEM results of displacement $x$ ($tol=10^{-10}$); the third column (c,g,k) shows the predictions of displacement $x$ by the pretraining phase of PreFine-Homo; and the fourth column (d,h,l) shows the predictions of displacement $x$ by the fine-tuning phase of PreFine-Homo. Rows 1 to 3 correspond to different porosities: 0.35, 0.25, and 0.15.}
	\label{fig:GRF_res}
\end{figure}

\Cref{fig:GRF_res}a,e,i show the geometry and topology generated by GRF with different porosities (0.35, 0.25, 0.15). \Cref{fig:GRF_res}c,g,k present the results of the pretraining phase of PreFine-Homo, compared to the reference solutions obtained from FEM ($tol=10^{-10}$) shown in \Cref{fig:GRF_res}b,f,j. It is evident that the pretraining phase of PreFine-Homo did not perform well in this case. This indicates that PreFine-Homo requires the training and testing data distributions to be similar to achieve optimal performance. This is a common phenomenon in machine learning, as it is essential to ensure that the data distributions of the training and testing sets are aligned.

However, the subsequent fine-tuning phase of PreFine-Homo refines the results of the pretraining phase. \Cref{fig:GRF_res}d,h,l demonstrate that the fine-tuning phase of PreFine-Homo can effectively extrapolate to data not present in the training set. The results are in complete agreement with the reference solutions, as the fine-tuning phase employs the same iterative methods used in traditional FEM calculations. This ensures that PreFine-Homo possesses unlimited extrapolation capabilities.

\begin{figure}
	\centering
	\includegraphics[scale=0.65]{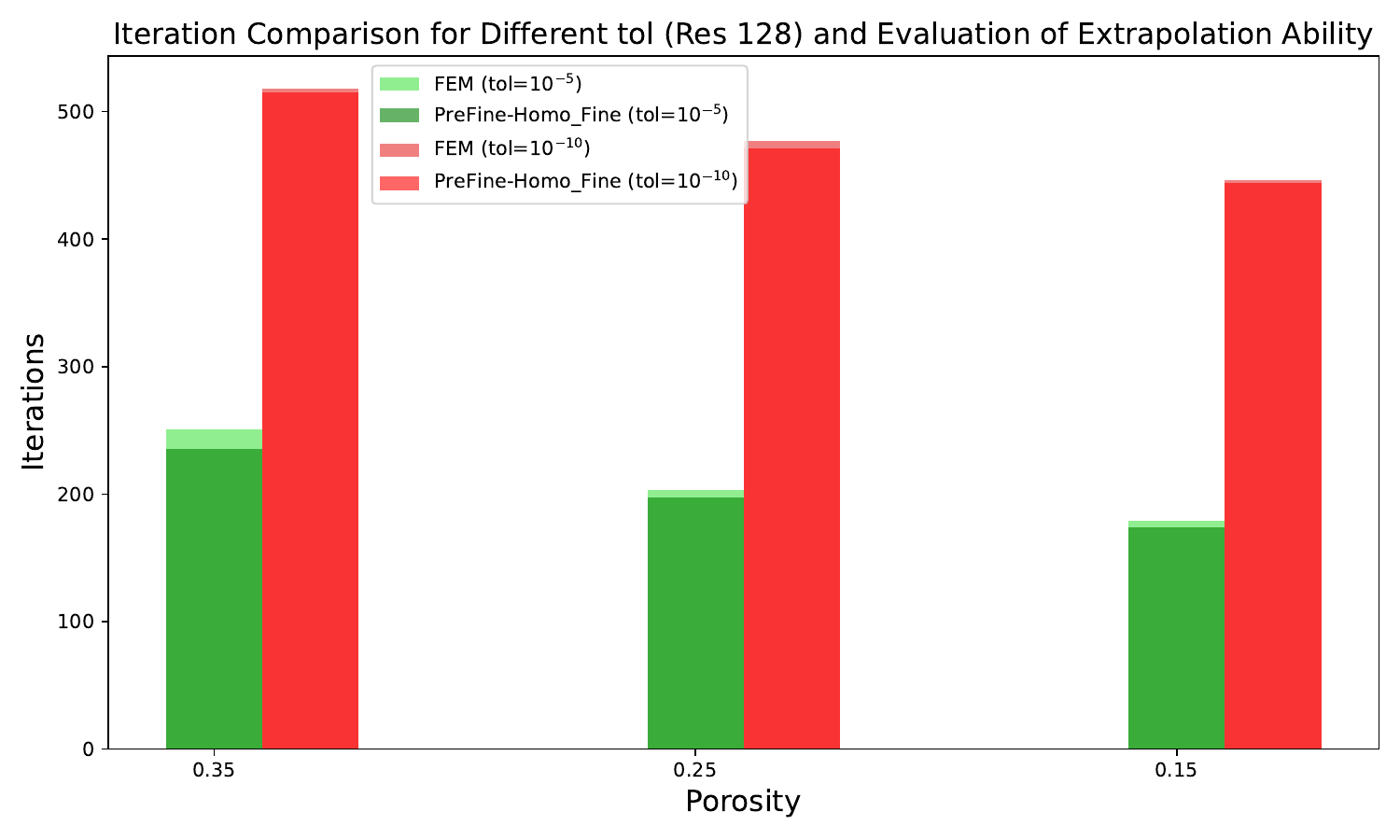}
	\caption{Comparison of the number of iterations required between the traditional PDE solver FEM and the fine-tuning phase of PreFine-Homo across different tolerances ($tol$) and porosities.}
	\label{fig:GRF_iterations}
\end{figure}

Next, we demonstrate the efficiency of PreFine-Homo on extrapolated datasets. \Cref{tab:PreFine-Homo_GRF} shows that the benefits of the fine-tuning phase of PreFine-Homo are significantly reduced. \Cref{fig:GRF_iterations} compares the number of iterations required in the fine-tuning phase of PreFine-Homo with those of traditional FEM. Although the benefits are modest, they still exist, indicating that even though GRF data was not included in the training set, the pretraining phase of PreFine-Homo exhibits some extrapolation capability. This suggests that there is a universal connection between different data distributions, albeit with varying degrees of strength.

\begin{table}
	\caption{The efficiency of PreFine-Homo on an average over all different porosity test cases with a resolution of 128. $tol$ is the convergence threshold for the iterative algorithm. FEM represents the average time and average number of iterations per load case for a traditional PDE solver. Different porosities refer to structures generated by Gaussian Random Fields (GRF) with porosities of 0.35, 0.25, and 0.15. The pretraining model of PreFine-Homo was trained on TPMS data and did not include GRF data. PreFine-Homo\_Pre denotes the average time for testing a single case in the pretraining phase of PreFine-Homo. PreFine-Homo\_Fine represents the average time and average number of iterations per load case for testing a single case in the fine-tuning phase of PreFine-Homo.}
	\label{tab:PreFine-Homo_GRF}
	\centering
	\begin{adjustbox}{max width=\textwidth}
		\begin{tabular}{cccccccc}
			\toprule 
			& \multicolumn{3}{c}{$tol=10^{-5}$} &  & \multicolumn{3}{c}{$tol=10^{-10}$} \\
			\cmidrule{2-4} \cmidrule{6-8} 
			Porosity & 0.35 & 0.25 & 0.15 &  & 0.35 & 0.25 & 0.15 \\
			\midrule
			FEM: sec/iters & 2645/250.8 & 3197/203.5 & 2522/179.0 &  & 5295/517.8 & 5893/477.0 & 5939/446.0 \\
			\midrule 
			PreFine-Homo\_Pre: sec & 0.4251 & 0.4362 & 0.4189 &  & 0.4251 & 0.4362 & 0.4189 \\
			\midrule 
			PreFine-Homo\_Fine: sec/iters & \textbf{2460/235.2} & \textbf{2850/197.2} & \textbf{2427/174.0} &  & \textbf{5278/515.3} & \textbf{5576/471.3} & \textbf{5826/444.3} \\
			\bottomrule
		\end{tabular}
	\end{adjustbox}
\end{table}

In summary, the pretraining phase of PreFine-Homo exhibits limited extrapolation capabilities, while the fine-tuning phase provides unlimited generalization capabilities, as it employs traditional FEM iterative algorithms. This means that PreFine-Homo can generalize to any geometry and material. The fine-tuned data can be used to retrain the pretraining model, as discussed in detail in \Cref{sec:unlimited_gener}.

\section{Discussion} \label{sec:Discussion}

\subsection{All Roads Lead to Rome}
\begin{figure}
	\centering
	\includegraphics[scale=0.75]{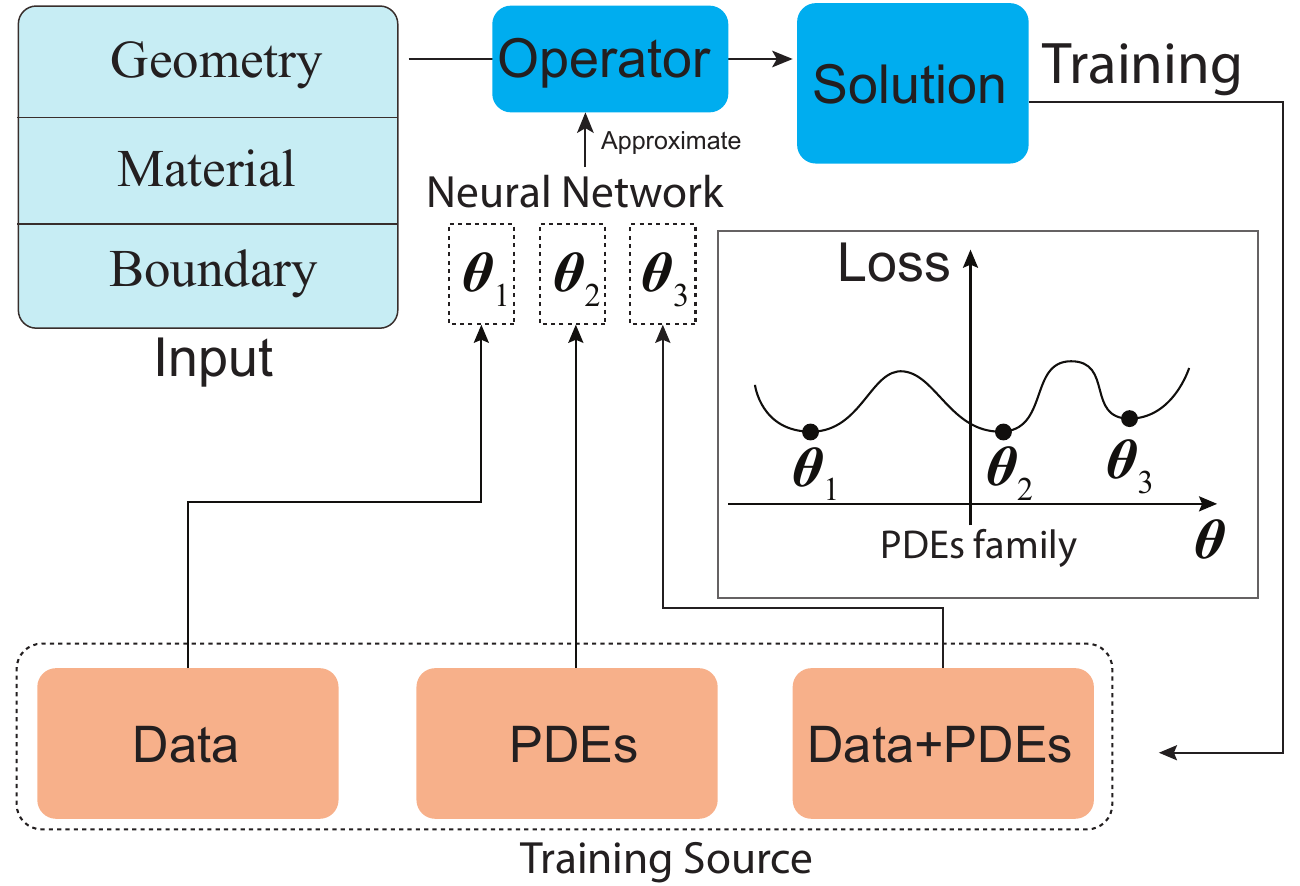}
	\caption{Illustration of training PDEs family operators using neural networks: $\boldsymbol{\theta}_{1}$ represents the parameters of a neural network trained purely through data-driven methods; $\boldsymbol{\theta}_{2}$ represents the parameters of a neural network trained purely through physical equations (PDEs); and $\boldsymbol{\theta}_{3}$ represents the parameters of a neural network trained by combining data and physical equations. \label{fig:PDEs-family}}
\end{figure}

The essence of PreFine-Homo lies in learning the operator mapping from the input function space (geometry and material properties) to the displacement solution space. Theoretically, training such an operator mapping from the input function space to the solution space can be achieved in multiple ways, as illustrated in \Cref{fig:PDEs-family}. For example, the Fourier Neural Operator (FNO) successfully trains PDE family operators purely through data \cite{li2020fourier}. Similarly, the Physics-Informed Neural Operator (PINO)  \cite{li2024physics} and Variational Neural Operator (VINO) \cite{eshaghi2025variational} can train PDEs family operators solely based on physical equations. Finally, PINO and VINO can also train PDEs family operators by combining data and physical equations. This demonstrates that there are multiple pathways to achieve the training of PDEs family operators, embodying the idea that "All Roads Lead to Rome."

Combining data and physical equations offers more advantages compared to using only data or only physical equations. Pure data-driven training requires a significantly large amount of data, while training solely based on physical equations has been shown to make neural network optimization more challenging \cite{li2020fourier}. On the other hand, combining data and physical equations can compensate for insufficient training data through the physical equations, while also leveraging data to make operator optimization easier. Therefore, in theory, the combination of data and physical equations is the optimal approach for training PDEs family operators. However, this approach may face challenges in balancing the trade-off between data and physical equation losses, which could require additional hyperparameter tuning.

The idea of "All Roads Lead to Rome" is analogous to the notion that there is not just one way to achieve intelligence. For instance, studies have shown that modifying just one pixel in an image can completely disrupt the performance of computer vision algorithms in classification tasks \cite{su2019one}, whereas the human brain remains unaffected by such minor changes. This suggests that while both the human brain and artificial intelligence can achieve excellent results in computer vision tasks, the mechanisms by which they achieve intelligence are fundamentally different. Similarly, in the context of AI for PDEs, there is more than one way to learn the PDEs family operators. Pure data-driven methods, pure physics-driven methods, and the combination of data and physics are all viable approaches to learning the underlying operators of PDEs families. This is an area worthy of future research to explore the advantages and disadvantages of each approach in more details.

\subsection{Benefits of the Fine-Tuning Phase in PreFine-Homo\label{sec:benifits_fine_tuning}}

It is easy to observe that the number of iterations required in the fine-tuning phase of PreFine-Homo is significantly reduced compared to traditional numerical algorithms. Below, we provide a convergence analysis of iterative methods for linear systems to explain this phenomenon.

Consider the iterative equation:
\begin{equation}
	\boldsymbol{X}^{(k)} = \boldsymbol{B}\boldsymbol{X}^{(k-1)} + \boldsymbol{f}. \label{eq:iterative_way}
\end{equation}
Clearly, the exact solution $\boldsymbol{X}^{*}$ satisfies:
\begin{equation}
	\boldsymbol{X}^{*} = \boldsymbol{B}\boldsymbol{X}^{*} + \boldsymbol{f}. \label{eq:exact_solution}
\end{equation}
We define the error as:
\begin{equation}
	\boldsymbol{e}^{(k)} = \boldsymbol{X}^{(k)} - \boldsymbol{X}^{*}. \label{eq:iter_error}
\end{equation}
Substituting \Cref{eq:iterative_way} and \Cref{eq:exact_solution} into \Cref{eq:iter_error}, we obtain:
\begin{equation}
	\begin{aligned}
		\boldsymbol{e}^{(k)} &= (\boldsymbol{B}\boldsymbol{X}^{(k-1)} + \boldsymbol{f}) - (\boldsymbol{B}\boldsymbol{X}^{*} + \boldsymbol{f}) \\
		&= \boldsymbol{B}(\boldsymbol{X}^{(k-1)} - \boldsymbol{X}^{*}) = \boldsymbol{B}\boldsymbol{e}^{(k-1)}.
	\end{aligned}
\end{equation}
By recursion, we have:
\begin{equation}
	\boldsymbol{e}^{(k)} = \boldsymbol{B}^{k}\boldsymbol{e}^{(0)}.
\end{equation}
Note that the quality of the initial solution is mathematically equivalent to the difference in $\boldsymbol{e}^{(0)}$, i.e., $\boldsymbol{e}^{(0)} = \boldsymbol{X}^{(0)} - \boldsymbol{X}^{*}$. In other words, the better the initial solution, the smaller the norm $\|\boldsymbol{e}^{(0)}\|$. Using the definition of the subordinate matrix norm, we obtain:
\begin{equation}
	\|\boldsymbol{B}^{k}\| = \max_{\boldsymbol{e}^{(0)}} \frac{\|\boldsymbol{B}^{k}\boldsymbol{e}^{(0)}\|}{\|\boldsymbol{e}^{(0)}\|} \geq \frac{\|\boldsymbol{B}^{k}\boldsymbol{e}^{(0)}\|}{\|\boldsymbol{e}^{(0)}\|}.
\end{equation}
\begin{equation}
	\|\boldsymbol{e}^{(k)}\| = \|\boldsymbol{B}^{k}\boldsymbol{e}^{(0)}\| \leq \|\boldsymbol{B}^{k}\| \cdot \|\boldsymbol{e}^{(0)}\|. \label{eq:iter_error_evolution}
\end{equation}
This clearly implies that, to achieve the same accuracy $\|\boldsymbol{e}^{(k)}\|$, a better initial solution reduces the upper bound on the number of iterations $k$ required. This theoretically explains why the better initial solution provided by the pretraining phase of PreFine-Homo reduces the number of iterations in the fine-tuning phase, especially in cases where traditional PDEs solvers fail to provide a good initial guess.

It is also evident that the benefits of the fine-tuning phase of PreFine-Homo diminish as the convergence threshold ($tol$) decreases, as shown in \Cref{fig:geo_iterations} and \Cref{fig:mater_iterations}. By transforming \Cref{eq:iter_error_evolution}, we obtain:
\begin{equation}
	\frac{\|\boldsymbol{e}^{(k)}\|}{\|\boldsymbol{e}^{(0)}\|} \leq \|\boldsymbol{B}^{k}\|. \label{eq:ratio_error}
\end{equation}
Let $\varepsilon = \|\boldsymbol{B}^{k}\|$. Then, \Cref{eq:ratio_error} can be rewritten as:
\begin{equation}
	\frac{\|\boldsymbol{e}^{(k)}\|}{\|\boldsymbol{e}^{(0)}\|} \leq \varepsilon. \label{eq:elison}
\end{equation}
We transform $\|\boldsymbol{B}^{k}\| = \varepsilon$ as follows:
\begin{align}
	\|\boldsymbol{B}^{k}\|^{\frac{1}{k}} &= \varepsilon^{\frac{1}{k}} \nonumber \\
	\ln\|\boldsymbol{B}^{k}\|^{\frac{1}{k}} &= \ln\varepsilon^{\frac{1}{k}} \nonumber \\
	k &= \frac{-\ln\varepsilon}{-\ln\|\boldsymbol{B}^{k}\|^{\frac{1}{k}}}. \label{eq:iter_num_discussion}
\end{align}
The asymptotic convergence rate is defined as:
\begin{equation}
	R(\boldsymbol{B}) = \lim_{k \rightarrow +\infty} -\ln\|\boldsymbol{B}^{k}\|^{\frac{1}{k}}. \label{eq:liter_ratio}
\end{equation}
The asymptotic convergence rate depends only on the iteration matrix $\boldsymbol{B}$ and is independent of the number of iterations $k$ and the initial solution. For large $k$, substituting \Cref{eq:liter_ratio} into \Cref{eq:iter_num_discussion}, we obtain:
\begin{equation}
	k = \frac{-\ln\varepsilon}{R(\boldsymbol{B})}. \label{eq:K_elison}
\end{equation}

The difference in the convergence threshold ($tol$) essentially corresponds to the difference in $\|\boldsymbol{e}^{(k)}\|$. Let $tol = 10^{-m}$. Then, $\|\boldsymbol{e}^{(k)}\|$ can be approximated as equal to $tol$. Thus, \Cref{eq:elison} becomes:
\begin{align}
	\frac{10^{-m}}{\|\boldsymbol{e}^{(0)}\|} &\leq \varepsilon \nonumber \\
	\ln\frac{10^{-m}}{\|\boldsymbol{e}^{(0)}\|} &\leq \ln\varepsilon \nonumber \\
	-m\ln10 - \ln\|\boldsymbol{e}^{(0)}\| &\leq \ln\varepsilon. \label{eq:elison_tol}
\end{align}
Substituting \Cref{eq:elison_tol} into \Cref{eq:K_elison}, we obtain:
\begin{align}
	k &= \frac{-\ln\varepsilon}{R(\boldsymbol{B})}  \nonumber \\
	&\leq \frac{m\ln10 + \ln\|\boldsymbol{e}^{(0)}\|}{R(\boldsymbol{B})} = \frac{m\ln10}{R(\boldsymbol{B})} + \frac{\ln\|\boldsymbol{e}^{(0)}\|}{R(\boldsymbol{B})} \label{eq:k_R}.
\end{align}

From \Cref{eq:k_R}, we observe that $m\ln10/R(\boldsymbol{B})$ represents the influence of the convergence threshold ($tol$) on the number of iterations, while $\ln\|\boldsymbol{e}^{(0)}\|/R(\boldsymbol{B})$ represents the influence of the initial solution on the number of iterations. Clearly, \Cref{eq:k_R} shows that as the convergence threshold decreases ($m$ increases), the impact of the initial solution on the number of iterations diminishes. This explains why the benefits of the fine-tuning phase of PreFine-Homo decrease as the convergence threshold ($tol$) decreases, as shown in \Cref{fig:geo_iterations} and \Cref{fig:mater_iterations}.

\subsection{A Self-Learning Closed-Loop Framework Based on Transfer Learning\label{sec:self_learning}}

Considering the high-precision numerical solutions obtained after the fine-tuning phase of PreFine-Homo, we can utilize these solutions to retrain the pretraining phase of PreFine-Homo.

\begin{figure}
	\centering
	\includegraphics[scale=0.75]{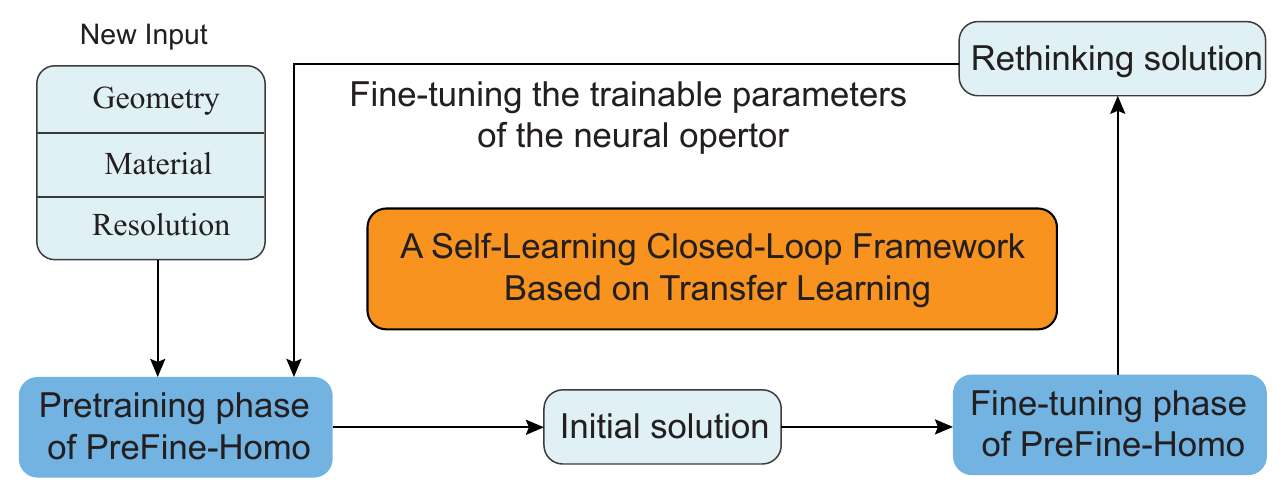}
	\caption{Self-Learning Closed-Loop Framework Based on Transfer Learning: Use transfer learning to reuse the solution of the PreFine-Homo fine-tuning phase to train the PreFine-Homo pre-trained model. \label{fig:transfer}}
\end{figure}

Given the limited amount of fine-tuned data, the cost of retraining would be significant. Thus, it is necessary to leverage transfer learning strategies to effectively utilize these high-precision fine-tuned solutions. There are various transfer learning strategies, and we will adopt the most common parameter-based fine-tuning approach in the future. Parameter-based fine-tuning can be implemented in two ways: full fine-tuning and lightweight fine-tuning. Recently, a widely used fine-tuning technique in large models is Low-Rank Adaptation (LoRA). While there are already results demonstrating the effectiveness of various transfer learning strategies in Physics-Informed Neural Networks (PINNs) \cite{wang2025transfer}, we believe that transfer learning can also be applied to operator learning in the future, enabling efficient fine-tuning based on small datasets.

It is worth emphasizing that the fine-tuning phase of PreFine-Homo ensures the model's unlimited extrapolation capabilities as shown in \Cref{sec:Extra}. Therefore, in theory, PreFine-Homo can achieve high-precision results on any dataset. These results can be used to self-learn and retrain the pretrained model of PreFine-Homo, forming a closed self-learning loop, as illustrated in \Cref{fig:transfer}. As a result, the model possesses unlimited growth potential. As the amount of data increases, the initial solutions provided by the pretrained model of PreFine-Homo will become increasingly accurate.

\subsection{Unlimited Generalization and Self-Learning Capabilities of PreFine-Homo\label{sec:unlimited_gener}}

PreFine-Homo possesses unlimited generalization capabilities because, during the fine-tuning phase, we employ traditional FEM iterative algorithms. This means that PreFine-Homo can generalize to any geometry and material. Although, as shown in \Cref{fig:GRF_res}, there is a gap between the predictions of PreFine-Homo in the pretraining phase and the exact FEM reference solutions, the model's predictions as initial solutions are still better than random initial guesses. This leads to a reduction in the number of iterations required for convergence.

In summary, the higher the accuracy of PreFine-Homo in the pretraining phase, the fewer iterations are needed in the fine-tuning phase. This implies that as the amount of data increases in the future, and PreFine-Homo encounters more diverse geometries and materials, it can continuously improve and become faster. This is a fundamental difference between PreFine-Homo and traditional FEM homogenization algorithms. While FEM algorithms lack the ability to grow and improve, our proposed PreFine-Homo framework possesses self-learning capabilities, enabling it to evolve and become more efficient with more and more data.

\section{Conclusion} \label{sec:Conclusion}

In this paper, we introduce a pretraining-finetuning framework for homogenization, named PreFine-Homo. In the pretraining phase, a neural operator is trained using a large dataset, where we adopt the Fourier Neural Operator (FNO) as the neural operator. The fine-tuning phase utilizes a traditional iterative algorithm, where the solution predicted by the pretraining phase serves as the initial solution for iteration.
The results show that PreFine-Homo demonstrates exceptional performance across various geometries, materials, and resolutions. We have validated the model using TPMS structures. Compared to traditional finite element homogenization programs, PreFine-Homo offers an efficiency improvement of nearly 1,000 times in predicting the required displacement, while maintaining accuracy close to that of finite element reference solutions. Therefore, it possesses both superior efficiency and high accuracy. Additionally, by integrating data across different resolutions, PreFine-Homo can utilize data of any resolution. This is particularly important in practical scenarios, as data often comes with varying resolutions.
Due to the fine-tuning phase, PreFine-Homo possesses unlimited extrapolation capability. In the future, as the amount of data increases, PreFine-Homo's generalization ability will become even stronger. Therefore, PreFine-Homo is a model capable of continuous learning.

However, PreFine-Homo currently has certain limitations. For instance, we have only considered linear problems thus far. Nevertheless, we believe that this framework is still applicable to nonlinear problems. Given that nonlinear computations have always been a challenge in computational mechanics and considering the powerful nonlinear fitting capability of neural networks, we believe PreFine-Homo has significant potential for addressing nonlinear problems. Therefore, we will extend PreFine-Homo to nonlinear problems in the future, such as hyperelasticity, elastoplasticity with history-dependent loading, and viscoelasticity with rate-dependent loading.
At present, the pretraining phase of PreFine-Homo is entirely data-driven. However, recent studies have shown that it is possible to train operators purely based on physical equations \citep{eshaghi2025variational, li2024physics}. Hence, future research could explore the physics-informed neural operator  to reduce the dependence on large datasets.
Additionally, errors in low-resolution data are generally larger than those in high-resolution data. Therefore, ensuring data reliability is crucial. This highlights a promising research direction on how to effectively train models with data that may contain errors.
During training and testing, we set the model's elastic modulus to 1 and then linearly scale it by scaling factor, assuming that the material is homogeneous. However, in the case of non-homogeneous materials, this approach is not feasible.

Although the model has not yet been validated for numerical homogenization in other physical problems, such as thermal conductivity, we focused on more challenging issues like the effective elastic tensor in this paper. Using this approach to study the prediction of effective properties for other physical problems presents an exciting research direction for the future.

\bmsection*{Author contributions}

Yizheng Wang designed the PreFine-Homo model, performed the numerical experiment, and wrote the paper. Xiang Li designed the traditional numerical algorithm, computed the data, and wrote the paper.  Ziming Yan designed the traditional numerical algorithm and revised the paper. Shuaifeng Ma desighed the PDEs solver based on FFT. Bokai Liu, and Jinshuai Bai reviewed the paper.  
Xiaoying Zhuang supervised the project.
Timon Rabczuk reviewed the paper and supervised the project. Yinghua Liu supervised the project and provided supercomputer computing resources.

\bmsection*{Acknowledgement}
This work is supported by the National Natural Science Foundation of China (No.12332005 and No.12162012) and  
Hainan Provincial Natural Science Foundation of China (No.121RC536). The authors would like to thank  Yuqing Du for revising the picture, Zhiqiang Meng and Yang Liu for helpful discussions. Yizheng Wang acknowledges the scholarship support from the Tsinghua University Future Scholars Program and the scholarship from Bauhaus University in Weimar, Germany.

\bmsection*{Financial disclosure}

None reported.

\bmsection*{Conflict of interest}

The authors declare no potential conflict of interests.

\bibliography{bibtex_FNO_effective}

\appendix
	
\bmsection{Solve PDEs using FFT} \label{sec:FFT}
In theory, solving periodic problems using the Fast Fourier Transform (FFT) method is faster than using the Finite Element Method (FEM). However, applying the FFT-based method to porous structures is particularly challenging due to the presence of infinite phase contrast at the microscale, which leads to severe Fourier ringing artifacts \cite{ladecky2023optimal} and necessitates a specific iterative framework. 

In this study, we also employed the FFT-based method to compute the elastic properties of Triply Periodic Minimal Surface (TPMS) structures. The FFT-based method was implemented using the Augmented Lagrangian (AL) framework \cite{moulinec2014comparison} combined with Willot discretization \cite{willot2015fourier}. The results show that the computation time of the FFT-based method varies significantly depending on the structural complexity. More complex structures require a greater number of iterations, with computation times ranging from 500 seconds to as long as 100,000 seconds. In contrast, as shown in \Cref{tab:PreFine-Homo_geo_time}, the results obtained using FEM are relatively stable.

In summary, the FFT-based method generally improves efficiency for solving periodic structures compared to FEM, but its computation time exhibits high variability and lacks the stability of FEM. Additionally, FEM is more versatile and widely applicable. Therefore, in this study, we adopt FEM as the method for data generation and as the benchmark for comparison.

\bmsection{The computational time of direct methods and iterative methods} \label{sec:computational_time_iter} 

Solving PDEs, whether linear or nonlinear, ultimately requires solving linear systems of equations. The computational time $T$ required to solve a symmetric positive definite linear system $KU = F$ can be approximated as:
\begin{equation}
	T = C n^{\alpha},
\end{equation}
where $C$ and $\alpha$ depend on the method used to solve the linear system and the condition number and sparsity of the coefficient matrix $\boldsymbol{K}$. Here, $n$ is the total number of unknowns, which corresponds to the dimension (DOF: degrees of freedom) of the solution vector $\boldsymbol{U}$, and $\boldsymbol{F}$ is a given vector of the same dimension as $\boldsymbol{U}$. 

There are generally two types of algorithms for solving linear systems: direct methods and iterative methods \citep{finite_element_book}. Both approaches have their advantages and disadvantages. Direct methods provide a precise estimate of the computational cost in advance, whereas the number of operations required by iterative methods depends on the initial guess and the desired accuracy of the solution (threshold of the iterative method). Although the coefficient $C_{iter}$ of iterative methods is significantly higher than that of direct methods $C_{dir}$, the exponent $\alpha_{iter}$ of iterative methods is lower than that of direct methods $\alpha_{dir}$:
\begin{equation}
	\begin{aligned}T_{iter} & =C_{iter}n^{\alpha_{iter}}\\
		T_{dir} & =C_{dir}n^{\alpha_{dir}}
	\end{aligned}
\end{equation}

For large-scale problems, the benefit of $n^{\alpha}$ far outweighs the impact of $C$. Therefore, iterative methods are generally more computationally efficient than direct methods for large-scale problems \cite{zhang2024blending}. Additionally, iterative methods are more amenable to parallel computing compared to direct methods.

\bmsection{Introduction to iterative algorithms for linear systems} \label{sec:iterative_method} 
Below, we introduce iterative algorithms for solving linear systems. First, consider a two-dimensional Poisson equation:
\begin{equation}
	\nabla^{2} p = f,
\end{equation}
where $p$ is the field function to be solved, and $f$ is a given function. Here, we approximate the Laplacian operator using finite differences:
\begin{equation}
	\begin{aligned}\frac{P_{I+1,J}-2P_{I,J}+P_{I-1,J}}{h_{x}^{2}}+\\
		\frac{P_{I,J+1}-2P_{I,J}+P_{I,J-1}}{h_{y}^{2}} & =F_{I,J}
	\end{aligned}
	\label{eq:FD_possion}
\end{equation}
where $I$ and $J$ are grid point indices, with $I = 1, 2, \dots, N_{x}$ and $J = 1, 2, \dots, N_{y}$. Here, $h_{x}$ and $h_{y}$ are the grid sizes in the $x$ and $y$ directions, respectively, and we assume the grid is uniform in both directions. If $h_{x} = h_{y} = h$, we can rewrite \Cref{eq:FD_possion} in matrix form:
\begin{equation}
	(\boldsymbol{D} - \boldsymbol{L} - \boldsymbol{U})\boldsymbol{P} = h^{2}\boldsymbol{F}, \label{eq:matrix_DLU}
\end{equation}
where $\boldsymbol{D}$, $\boldsymbol{L}$, and $\boldsymbol{U}$ are the diagonal matrix, the negative of the lower triangular matrix, and the negative of the upper triangular matrix, respectively, derived from \Cref{eq:FD_possion}. Here, $\boldsymbol{P}$ and $\boldsymbol{F}$ are the discrete matrix forms of $p$ and $f$, respectively. Below, we introduce common iterative algorithms for solving linear systems, including the Jacobi method, the Gauss-Seidel method, the Successive Over-Relaxation (SOR) method, and Preconditioned Conjugate Gradient (PCG). It is worth noting that once an iterative algorithm is chosen, its convergence rate is fixed and independent of the initial guess. While the quality of the initial guess can affect the number of iterations required for convergence, it does not influence the convergence rate.

\bmsubsection{Jacobi Method}

We transform \Cref{eq:matrix_DLU} into an iterative algorithm:
\begin{equation}
	\boldsymbol{P}^{(k+1)} = \boldsymbol{D}^{-1}(\boldsymbol{L} + \boldsymbol{U})\boldsymbol{P}^{(k)} + h^{2}\boldsymbol{D}^{-1}\boldsymbol{F}. \label{eq:jacobi_iter}
\end{equation}

\Cref{eq:jacobi_iter} represents the classical Jacobi iteration method. The Jacobi method can be unstable, so the damped Jacobi method improves stability by introducing a damping factor $\alpha$:
\begin{equation}
	\begin{aligned}\boldsymbol{P}^{(k+1)}= & (1-\alpha)\boldsymbol{P}^{(k)}+\\
		& \alpha[\boldsymbol{D}^{-1}(\boldsymbol{L}+\boldsymbol{U})\boldsymbol{P}^{(k)}+h^{2}\boldsymbol{D}^{-1}\boldsymbol{F}]
	\end{aligned}
	\label{eq:jacobi_iter_damp}
\end{equation}

The Jacobi method is simple and parallelization-friendly, but its convergence rate is slow. Below, we introduce the Gauss-Seidel method, which improves the convergence rate.

\bmsubsection{Gauss-Seidel Method}

The Gauss-Seidel method replaces part of $\boldsymbol{P}^{(k)}$ in \Cref{eq:jacobi_iter} with the already computed results $\boldsymbol{P}^{(k+1)}$:
\begin{equation}
	\boldsymbol{P}^{(k+1)} = \boldsymbol{D}^{-1}\boldsymbol{L}\boldsymbol{P}^{(k+1)} + \boldsymbol{D}^{-1}\boldsymbol{U}\boldsymbol{P}^{(k)} + h^{2}\boldsymbol{D}^{-1}\boldsymbol{F}.
\end{equation}
Clearly, the Gauss-Seidel method is sequential and cannot be effectively parallelized. While it accelerates convergence, it sacrifices the computational efficiency gained from vectorization.

We can further improve the convergence rate by introducing the Successive Over-Relaxation (SOR) method.

\bmsubsection{Successive Over-Relaxation (SOR) Method}

The SOR method combines the results of the Gauss-Seidel method with the previous iteration results using a linear weighting:
\begin{equation}
	\boldsymbol{P}^{(k+1)} = w\boldsymbol{P}_{gs}^{(k+1)} + (1 - w)\boldsymbol{P}^{(k)},
\end{equation}
where $\boldsymbol{P}_{gs}^{(k+1)}$ is the result computed using the Gauss-Seidel method. The SOR method can significantly accelerate convergence when the relaxation parameter $w$ is chosen appropriately.

The Jacobi, Gauss-Seidel, and SOR methods often face two issues: (1) they may diverge when solving non-symmetric or non-positive definite systems, and (2) their convergence rate for low-frequency components is slow \citep{zhang2024blending}. This is because the Jacobi and Gauss-Seidel methods primarily relax high-frequency components, while their effectiveness for low-frequency components is limited. Some improved iterative algorithms, such as the Multigrid method \citep{briggs2000multigrid}, can address these issues by performing iterations simultaneously at multiple scales (different resolutions), effectively relaxing all frequency modes, especially low-frequency ones, thereby accelerating convergence.

To unify the Jacobi, Gauss-Seidel, and SOR methods, we consider the following iterative formula based on \Cref{eq:matrix_DLU}:
\begin{equation}
	\begin{aligned}\boldsymbol{P}^{(k+1)} & =\boldsymbol{P}^{(k)}+\boldsymbol{A}^{-1}\boldsymbol{R}^{(k)}\\
		\boldsymbol{A} & =\boldsymbol{D}-\boldsymbol{L}-\boldsymbol{U}\\
		\boldsymbol{R}^{(k)} & =h^{2}\boldsymbol{F}-\boldsymbol{A}\boldsymbol{P}^{(k)}
	\end{aligned}
\end{equation}
Since computing $\boldsymbol{A}^{-1}$ is computationally expensive, the essence of iterative methods lies in approximating $\boldsymbol{A}^{-1}$. Different approximations of $\boldsymbol{A}^{-1}$ give rise to the Jacobi, Gauss-Seidel, and SOR methods, as shown in \Cref{tab:Jacobi_GS}.

\begin{table}
	\caption{Jacobi, Gauss-Seidel, and SOR Methods: Different Approximations of $\boldsymbol{A}^{-1}$\label{tab:Jacobi_GS}}
	\centering
	\begin{tabular}{cc}
		\toprule 
		Method & Approximation of $\boldsymbol{A}^{-1}$\tabularnewline
		\midrule 
		Jacobi & $\boldsymbol{A}^{-1} = \boldsymbol{D}^{-1}$\tabularnewline
		Gauss-Seidel & $\boldsymbol{A}^{-1} = (\boldsymbol{D} - \boldsymbol{L})^{-1}$\tabularnewline
		SOR & $\boldsymbol{A}^{-1} = (\boldsymbol{D}/w - \boldsymbol{L})^{-1}$\tabularnewline
		\bottomrule
	\end{tabular}
\end{table}

\bmsubsection{Preconditioned Conjugate Gradient (PCG) Algorithm}
Since the 3D elastic problem involves large and sparse systems of symmetric positive definite matrices, our proposed PreFine-Homo framework employs the Preconditioned Conjugate Gradient (PCG) method as its iterative solver.
Several other iterative algorithms can operate on symmetric positive definite matrices, but PCG is the quickest and most reliable at solving large and sparse systems of symmetric positive definite matrices \citep{barrett1994templates}.

The PCG method is an enhancement of the classical Conjugate Gradient (CG) method, incorporating a preconditioner to accelerate convergence. The key idea behind PCG is to transform the original system of equations into an equivalent system with improved spectral properties, thereby reducing the number of iterations required to achieve a desired level of accuracy.

Given a linear system of the form:
\begin{equation}
	\boldsymbol{K}\boldsymbol{X} = \boldsymbol{f},
\end{equation}
where $\boldsymbol{K}$ is a symmetric positive definite matrix, $\boldsymbol{X}$ is the solution vector, and $\boldsymbol{f}$ is the right-hand side vector, the PCG algorithm can be summarized as follows:

\begin{enumerate}
	\item \textbf{Initialization:} Choose an initial guess $\boldsymbol{X}^{(0)}$ and compute the residual $\boldsymbol{r}^{(0)} = \boldsymbol{f} - \boldsymbol{K}\boldsymbol{X}^{(0)}$. Select a preconditioner $\boldsymbol{M}$ such that $\boldsymbol{M}$ approximates $\boldsymbol{K}^{-1}$ and is easy to invert.
	
	\item \textbf{Iteration:} For each iteration $k = 0, 1, 2, \dots$, perform the following steps:
	\begin{itemize}
		\item Solve the preconditioned system: $\boldsymbol{z}^{(k)} = \boldsymbol{M}^{-1}\boldsymbol{r}^{(k)}$.
		\item Compute the search direction $\boldsymbol{p}^{(k)}$ using the conjugate direction update.
		\item Update the solution vector: $\boldsymbol{X}^{(k+1)} = \boldsymbol{X}^{(k)} + \alpha_k \boldsymbol{p}^{(k)}$, where $\alpha_k$ is the step size computed to minimize the residual.
		\item Update the residual: $\boldsymbol{r}^{(k+1)} = \boldsymbol{r}^{(k)} - \alpha_k \boldsymbol{K}\boldsymbol{p}^{(k)}$.
		\item Check for convergence: If the residual norm $||\boldsymbol{r}^{(k+1)}||$ is below a specified tolerance, stop the iteration.
	\end{itemize}
\end{enumerate}

The PCG method is particularly effective when the preconditioner $\boldsymbol{M}$ is well-chosen, as it significantly reduces the condition number of the system, leading to faster convergence. Common choices for preconditioners include diagonal (Jacobi) preconditioning, incomplete Cholesky factorization, and multigrid methods. In the PreFine-Homo, we use the incomplete Cholesky factorization for  $\boldsymbol{M}$, because $\boldsymbol{K}$ in 3D elasticity problems is a sparse and symmetric positive definite matrix.

\bmsection{The generation of the TPMS} \label{sec:TPMS}

We introduce the mathematical model of the TPMS.
A minimal surface is characterized to have a zero value of mean curvature at any given point of the surface. TPMS is infinitely periodic in three-dimensional space. Level-set approximation is a commonly used approach to represent a minimal surface. The mathematical model of TPMS features level-set equations combined with a series of trigonometric functions, enabling TPMS flexibility to manipulate its topology and mechanical properties. Level-set equations are normally described by
\begin{equation}
	\phi(x, y, z) = c, 
\end{equation}
where $x$, $y$, and $z$ are the spatial coordinates. $\phi(x, y, z)$ on the left-hand side of the equation represents a diffusive field in the 3D space. $\phi=c$ denotes that the level-set value c segments the 3d space of a TPMS unit cell into two connected sub-domains, with each sub-domain representing solid material and void, respectively. The value of c can be used to indicate the volumetric portion of the void. The porosity of the TPMS unit cell can be hence manipulated by adjusting the c value. 
The iso-surface of TPMS means that an iso-value equals c. 

A broadly utilized level-set equation of TPMS is named Schwarz Primitive formatted by
\begin{equation}
	\cos(2\pi x) + \cos(2\pi y) + \cos(2\pi z) = c
\end{equation}
The above level-set equation denotes a zero-thickness shell-like topology, known as the minimal surface. We need to further generate load-bearing TPMS lattice material based on the minimal surfaces. Two types of TPMS lattices are formulated based on minimal surfaces. The first type is named “Solid-networks”, satisfying
\begin{equation}
	\phi(x, y, z) > c
\end{equation}
For "solid-networks" TPMS cells, the volume with the diffusive field value $\phi(x, y, z) > c$ is recognized as a solid material, while the rest of the unit cell is recognized as void or pore. The second type is named  "sheet-networks". The "sheet-networks" TPMS is governed by
\begin{equation}
	-c \leq \phi(x, y, z) \leq c
\end{equation}
As its name suggests, "sheet-networks" selects a thin layer of volume around the zero-thickness shell of $\phi(x, y, z)=c$, representing a sheet-like topology of the unit cell. The "solid-networks" and "sheet-networks" TPMS present distinguishable topology characteristics and elastic mechanical properties, offering more flexibility to its "topology-driven" capability.

Another key topological parameter for TPMS is volume fraction (or relative density). Volume fraction refers to the ratio between the volume of solid material and the volume of the entire TPMS unit cell. The volume fraction is directly related to the effective elastic tensor and stiffness of a TPMS unit cell and hereby affects the homogenization elastic matrix. 

In this study, we created a series of TPMS unit cells for training and verifying the proposed neural operator-based homogenization method. In this study, we selected 3 different TPMS level-set equations, including  "Schoen Gyroid", "Schwarz Diamond", and "Fischer Kosh S". The level-set equations for each unit cell type are listed:
\begin{equation}
	\begin{aligned}
		&Schoen Gyroid:\sin(2\pi x)\cos(2\pi y) + \sin(2\pi y)\cos(2\pi z) \\
		& \qquad \qquad \qquad + \sin(2\pi z)\cos(2\pi x) = c \\
		&Schwarz Diamond:\cos(2\pi x)\cos(2\pi y)\cos(2\pi z)\\
		& \qquad \qquad \qquad - \sin(2\pi x)\sin(2\pi y)\sin(2\pi z) = c \\
		&Fischer Kosh S:\cos(4\pi x) \sin(2\pi y) \cos(2\pi z) + \\
		& \qquad \qquad \qquad \cos(2\pi x) \cos(4\pi y) \sin(2\pi z) \\
		& \qquad \qquad \qquad + \sin(2\pi x) \cos(2\pi y) \cos(4\pi z) = c
		\label{eq:TPMStype}
	\end{aligned}
\end{equation}

\bmsection{Scaling factor for effective elastic tensor} \label{sec:Factor_technique}

We can calculate the scaling factor by comparing the FNO-predicted effective elastic tensor with the ground truth values in the training set:
\begin{equation}
	R_{ijkl} = \sum_{m=1}^{N_{train}} \frac{E_{ijkl}^{gt(m)}}{E_{ijkl}^{fno(m)}},
\end{equation}
where $E_{ijkl}^{gt(m)}$ and $E_{ijkl}^{fno(m)}$ are the ground truth and FNO-predicted effective elastic tensors, respectively. Here, we do not perform tensor operations on $ijkl$. $N_{train}$ is the total number of training samples used to determine the scaling factor, which can be adjusted based on specific requirements.

It is worth noting that the scaling factor can improve the accuracy of PreFine-Homo, but only during the pretraining phase. In the fine-tuning phase, the scaling factor is unnecessary because PreFine-Homo already achieves very high displacement accuracy after fine-tuning, eliminating the need for factor adjustments to the effective elastic tensor.

\bmsection{Supplementary code}
	The code of this work will be available at \url{https://github.com/yizheng-wang/Research-on-Solving-Partial-Differential-Equations-of-Solid-Mechanics-Based-on-PINN}. Note to editor and reviewers: the link above will be made public upon the publication of this manuscript. During the review period, the data and source code can be made available upon request to the corresponding author.

\end{document}